\pdfoutput=1
\documentclass[english,aps,preprint,prl]{revtex4}
\usepackage[T1]{fontenc}
\usepackage[latin9]{inputenc}
\usepackage{amsmath}
\usepackage{graphicx}
\usepackage{esint}

\makeatletter

\@ifundefined{textcolor}{}
{%
 \definecolor{BLACK}{gray}{0}
 \definecolor{WHITE}{gray}{1}
 \definecolor{RED}{rgb}{1,0,0}
 \definecolor{GREEN}{rgb}{0,1,0}
 \definecolor{BLUE}{rgb}{0,0,1}
 \definecolor{CYAN}{cmyk}{1,0,0,0}
 \definecolor{MAGENTA}{cmyk}{0,1,0,0}
 \definecolor{YELLOW}{cmyk}{0,0,1,0}
}

\makeatother

\usepackage{babel}
\begin{document}

\preprint{}

\title{Multiscale approach to equilibrating model polymer melts}

\author{Carsten Svaneborg}

\email{science@zqex.dk}

\homepage{http://www.zqex.dk/}

\selectlanguage{english}%

\affiliation{University of Southern Denmark, Campusvej 55, DK-5230 Odense M, Denmark}

\author{Hossein Ali Karimi-Varzaneh}

\email{ali.karimi@conti.de}

\selectlanguage{english}%

\affiliation{Continental, PO Box 169, D-30001 Hannover, Germany}

\author{Nils Hojdis}

\email{nils.hojdis@conti.de}

\selectlanguage{english}%

\affiliation{Continental, PO Box 169, D-30001 Hannover, Germany}

\author{Frank Fleck}

\email{frank.fleck@conti.de}

\selectlanguage{english}%

\affiliation{Continental, PO Box 169, D-30001 Hannover, Germany}

\author{Ralf Everaers}

\email{ralf.everaers@ens-lyon.fr}

\selectlanguage{english}%

\affiliation{Univ Lyon, ENS de Lyon, Univ Claude Bernard, CNRS, Laboratoire de
Physique and Centre Blaise Pascal, F-69342 Lyon, France }
\begin{abstract}
We present an effective and simple multiscale method for equilibrating
Kremer Grest model polymer melts of varying stiffness. In our approach,
we progressively equilibrate the melt structure above the tube scale,
inside the tube and finally at the monomeric scale. We make use of
models designed to be computationally effective at each scale. Density
fluctuations in the melt structure above the tube scale are minimized
through a Monte Carlo simulated annealing of a lattice polymer model.
Subsequently the melt structure below the tube scale is equilibrated
via the Rouse dynamics of a force-capped Kremer-Grest model that allows
chains to partially interpenetrate. Finally the Kremer-Grest force
field is introduced to freeze the topological state and enforce correct
monomer packing. We generate $15$ melts of $500$ chains of $10.000$
beads for varying chain stiffness as well as a number of melts with
$1.000$ chains of $15.000$ monomers. To validate the equilibration
process we study the time evolution of bulk, collective and single-chain
observables at the monomeric, mesoscopic and macroscopic length scales.
Extension of the present method to longer, branched or polydisperse
chains and/or larger system sizes is straight forward.
\end{abstract}
\maketitle

\section{introduction}

Computer simulations of polymer melts and networks allow unprecedented
insights into the relation between microscopic molecular structure
and macroscopic material properties such as the viscoelastic response
to deformation, see e.g. \citep{snijkers2015perspectives,padding2011systematic,li2013challenges,everaers2004rheology}.
Such simulation studies rely on very large model systems to reliably
estimate material properties, and an important obstacle is the generation
of large well equilibrated model systems for long entangled polymer
chains.

What do we mean by equilibrium in the case of a linear homopolymer
polymer melt? 1) Polymeric liquids have bulk moduli comparable to
that of water, and they are nearly incompressible. Hence in equilibrium,
we expect model melt states without significant density fluctuations.
2) Single chains in a melt adopt self-similar random walk statistics
because excluded volume interactions are screened at length scales
sufficiently large compared to monomeric scales. Hence in equilibrium,
we expect model states without significant deviations from random
walk statistics. 3) At mesoscopic scales, many polymer chains pervade
the same volume, such that chains are strongly topologically entangled.
This gives rise to the well known plateau modulus.\citep{DoiEdwards_86}
Hence in equilibrium, we also require model melts that achieve the
correct entanglement density. 4) Finally at the monomeric scale we
require the correct monomeric packing, such that we can expect to
run long stable simulations where topology is preserved. Taken together,
these constraints couple the conformations on scales that range from
monomeric to macroscopic length scales. This makes the problem of
making equilibrated model melts particular difficult, and the problem
is acerbated in the case of heteropolymers or for branched polymers
which are of significant industrial interest.

Brute force equilibration of model polymer materials is typically
not feasible. Polymer materials display dynamics over a huge range
of time scales. Even for polymers of moderate size, their largest
conformational relaxation times are many orders of magnitude beyond
that which is currently available via brute force simulation. Monomeric
motion takes place on pico second time scales, whereas conformational
relaxation times can easily reach up to macroscopic time scales. For
a long linear polymer chains the dominant relaxation mechanism is
reptation\citep{de1971reptation,de1976dynamics,klein1978evidence}
which gives rise to relaxation times $\tau\sim N^{3}$ where $N$
is the number of monomers.\citep{de1976dynamics} In the case of star
shaped polymers, reptation is not possible and the dominant relaxation
mechanism becomes contour length fluctuations\citep{zamponi2005contour},
in which case the relaxation times are $\tau\sim\exp(N_{arm})$, where
$N_{arm}$ is the number of monomers in an arm.\citep{pearson1984viscoelastic} 

To our knowledge, there are three major strategies for equilibrating
model polymer melts that address the challenges raised above: a) algorithms
that attempts to construct equilibrium model melts with the correct
large-scale single chain statistics, b) algorithms that utilize unphysical
Monte Carlo (MC) moves to accelerate the dynamics compared to brute
force Molecular Dynamics (MD), which simulates the real physical polymer
dynamics, c) algorithms using different models e.g. utilizing softer
potentials and a push-off process to allow chains to cross to accelerate
the relaxation process. In the present approach we combine all these
approaches, but before presenting our approach we review examples
of these strategies found in the literature.

It is easy to generate single chain conformations with the desired
large scale chain statistics. Equilibration procedures following this
approach typically place the resulting chains randomly into the simulation
domain. However, when monomer packing is introduced, the presence
of density fluctuations in the initial state cause significant local
chain stretching and compression. Brown et al. \citep{BrownJCP1990}
were the first to recognize the importance of such density fluctuations.
This was analyzed in detail by Auhl et al.\citep{auhl2003equilibration},
who made two proposals of how to resolved the density fluctuations,
either to accelerate the relaxation utilizing double bridging moves,
or to pre-pack the chains in space to avoid density fluctuations.
This was done using Monte Carlo simulated annealing and accepting
only moves that reduce density fluctuations\citep{auhl2003equilibration} 

An completely different approach is has been proposed by Gao\citep{gao1995efficient}.
The idea is to start by an equilibrated liquid of monomers and then
to create bonds between the monomers corresponding to a melt of polymers.
This completely side steps the issue of density fluctuations, since
the monomeric liquid is also incompressible. However, the problem
becomes how to identify a set of potential bonds to that correspond
to a mono-disperse melt of long linear or branched polymers. To reach
near complete conversion Gao had to increase the search distance for
the last bonds, and to remove the last monomers that could not be
bonded. Whereas Gao performed instantaneous bonding on a frozen monomer
liquid, Barrat and coworkers\citep{perez2008polymer} extended the
method by allowing the monomers to move during bonding. This has the
effect of enhancing the search distance for bonding. This method still
has issues with producing mono-disperse melts, Barrat and coworkers
solved the problem by aborting the bonding procedure when 80\% of
the monomers are linked into mono-disperse chains, and then removing
the last 20\% of monomers. The resulting states were then compressed
to the target pressure, which globally deforms the chain statistics. 

Monte Carlo techniques (MC) have the advantage that unphysical moves
can be used to accelerate the relaxation dynamics compared to MD techniques,
which follow the physical dynamics. A key contribution to the equilibration
of polymer melts were the complex MC moves developed by Theodorou
and co-workers.\citep{karayiannis2002novel,mavrantzas1999end} End-bridging
moves works by identifying a the end monomer of one chain and an internal
monomer of another chain where the two monomers are in close proximity.
The move is performed by cutting the chain at the internal monomer
and attaching it to the end of the other chain.\citep{pant1995variable,mavrantzas1999end}
Double bridging moves works by identifying two pairs of bonded monomers
in spatial proximity. The move is performed by replacing the two intramolecular
bonds by two intermolecular bonds. The result of these moves is a
melt conformation with a new chemical connectivity. Compared with
end-bridging moves double-bridging preserves the chain length when
equivalent bead pairs along the polymer contours are chosen.

The double bridging moves are the best way currently known to accelerate
the polymer dynamics, but method suffers from two major problems:
1) as the chain length is increased the density of potential sites
for double bridging drops, and 2) the new proposed connectivity can
have a high configurational energy, hence necessitating further tricks
to relax the conformation to ensure a reasonable acceptance rate.
For instance, it was proposed to reconnect not just monomers, but
to grow small bridge segments in order to reduce the conformational
energy of the proposed new state.\citep{karayiannis2002novel} These
methods have been used to equilibrate linear melts of polyethylene
up to $1000$ monomers\citep{karayiannis2002novel,karayiannis2002atomistic},
and polydisperse polyethylene melts up to $5000$ monomers.\citep{uhlherr2002atomic}
They have also been applied to branched molecules\citep{karayiannis2003advanced,peristeras2005structure,ramos2007monte}
and grafted polymers\citep{daoulas2002detailed,daoulas2003variable}.

The first multi-scale approach was introduced by Subramanian\citep{subramanian2010topology,subramanian2011iterative},
who applied it to linear and branched melts. His idea was to start
by equilibrating a coarse representation of the polymer, and successively
rescale the simulation domain by while doubling the number of beads
in the polymer model. In this way polymer conformations are successively
equilibrated on smaller and smaller length scales. A more sophisticated
version of hierarchical equilibration has been studied by Zhang et
al.\citep{zhang2014equilibration}, where a range of blob chain models
were successively fine grained with force field that depended on the
scale of fine graining. The most recently proposed equilibration method
is that of Moreira et al.\citep{moreira2015direct}, who develop the
Auhl method further by applying a warm-up procedure where pair-interactions
are slowly introduced via a cap on the maximal force as well as the
cut-off distance of the pair interactions that is progressively raised
using an elaborate feed-back control mechanism during the equilibration
process. 

Equilibrated melts of atomistic polymer models can be obtained via
fine-graining from a coarse-grained polymer model. Theodorou and Suter\citep{theodorou1985detailed,theodorou1986atomistic}
studied polymer melts with atomistic models which they prepared by
growing atomistic polymer models bond-by-bond in the simulation domain
using a metropolis acceptance criterion while taking non-bonded interactions
into account when choosing bond angles. The resulting states were
then energy minimized. Carbone et al. \citep{carbone2010fine} produce
atomistic polymer melts by generating continuous (non-packed) random
walks and fine graining them using an atomistic polymer models. For
each continuous random walk, a corresponding atomistic polymer chain
is created by confining the configuration to follow the continuous
random walk, and intra-chain monomeric packing is slowly introduced
through a simulation with a soft push-off potential. In a second step,
the atomistic polymer chains are placed in the simulation domain,
and a second push-off procedure is performed to introduce inter-chain
monomeric packing. A similar approach was used by Kotelyanskii et
al. but using self-avoiding random walks on a cubic lattice for the
initial random walks, which resolves the packing problem.\citep{kotelyanskii1996building}
Recently, Sliozberg et al. equilibrated a one million atom system
of polyethylene using an united atom model. \citep{sliozberg2016fast}.
Similar to Theodorou and Suter the polymers are grown in the simulation
domain, taking chemical structure into account to some extend. The
resulting melt conformations are then simulated with a soft DPD inspired
potential to gentle introduce excluded volume interactions, until
they can be switched to the final united-atom force field. 

In the present paper, our aim is to present a new general, simple,
and computationally effective method of rapidly generating very large
equilibrated melts of polymers. We illustrate the method by creating
equilibrated monodisperse linear Kremer-Grest (KG)\citep{grest1986molecular}
polymer models. This polymer model is the standard model for Molecular
Dynamics simulations of polymers. The KG model is generic and describes
universal polymer properties without attempting to model chemical
details of specific polymer species. Chemical details can be introduced
in the KG model by varying the effective chain stiffness, which allows
us to use this model for studying universal properties of specific
polymer types.\citep{svaneborgContiII} Here we follow Auhl et al.\citep{auhl2003equilibration}
and study how to produce equilibrated melts for a wide range of chain
stiffnesses. The typical size of the melts we generate in this study
comprise $5-15\times10^{6}$ beads for chains of $15.000$ beads per
chain or $200$ entanglements per chain. These numbers are chosen
be about a factor of five above the state of the art e.g. \citep{zhang2014equilibration,moreira2015direct}.
However, we are by no means pushing the limitations of the present
equilibration approach.

\begin{figure}
\includegraphics[width=0.45\columnwidth]{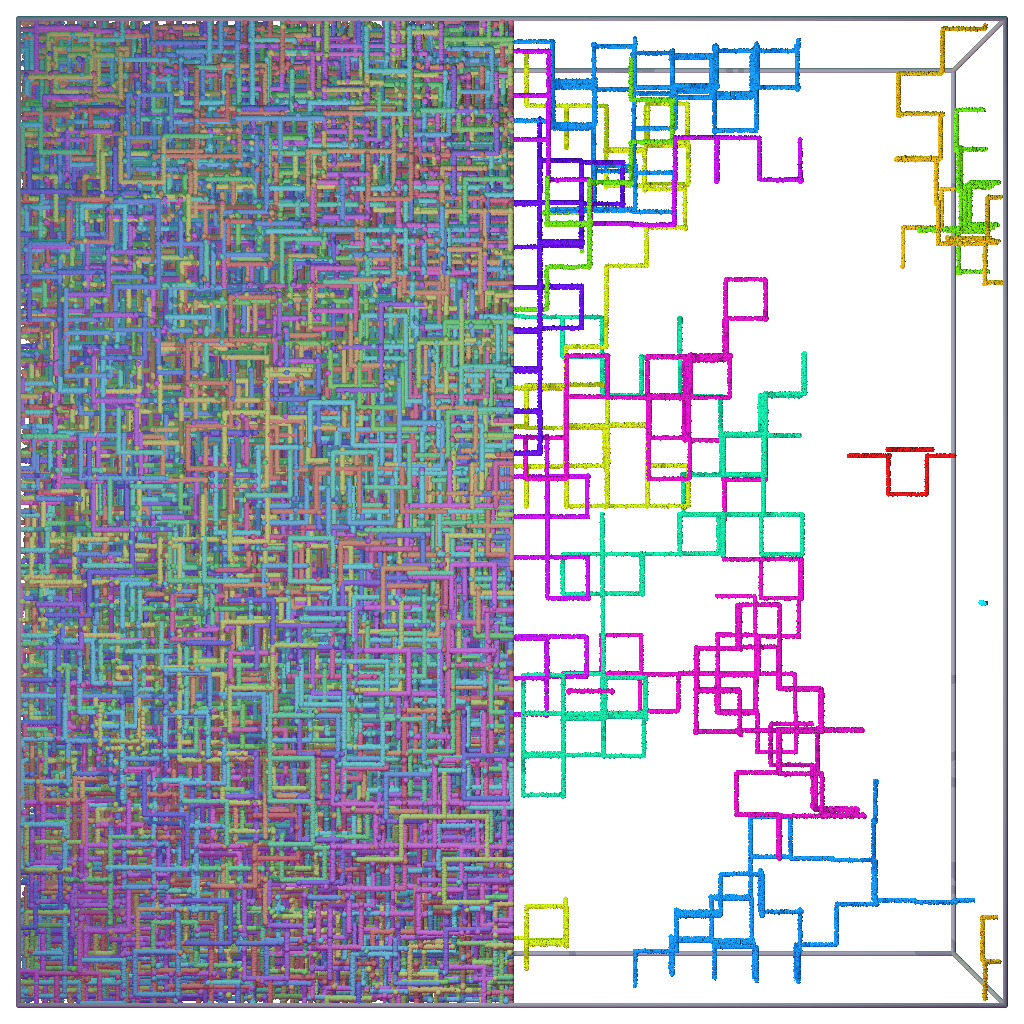}%
\includegraphics[width=0.45\columnwidth]{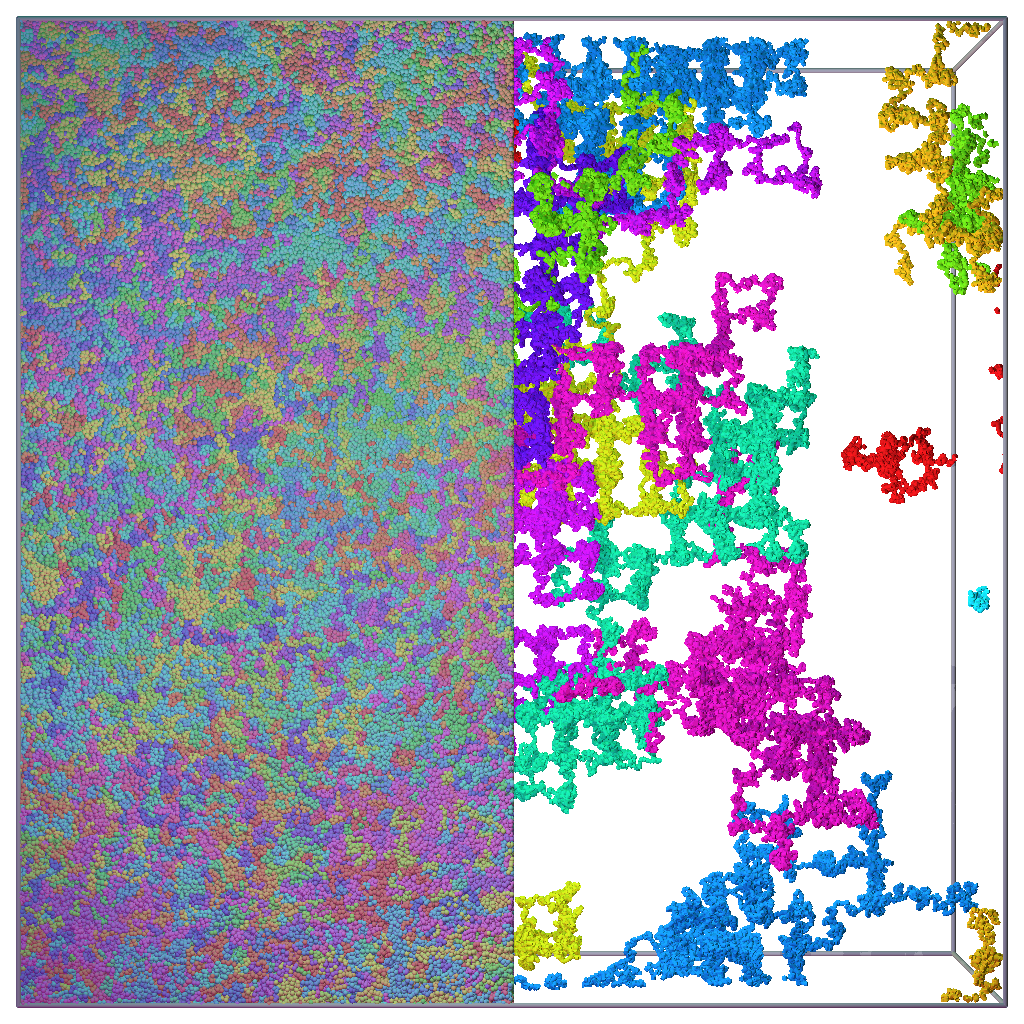}

\includegraphics[width=0.45\columnwidth]{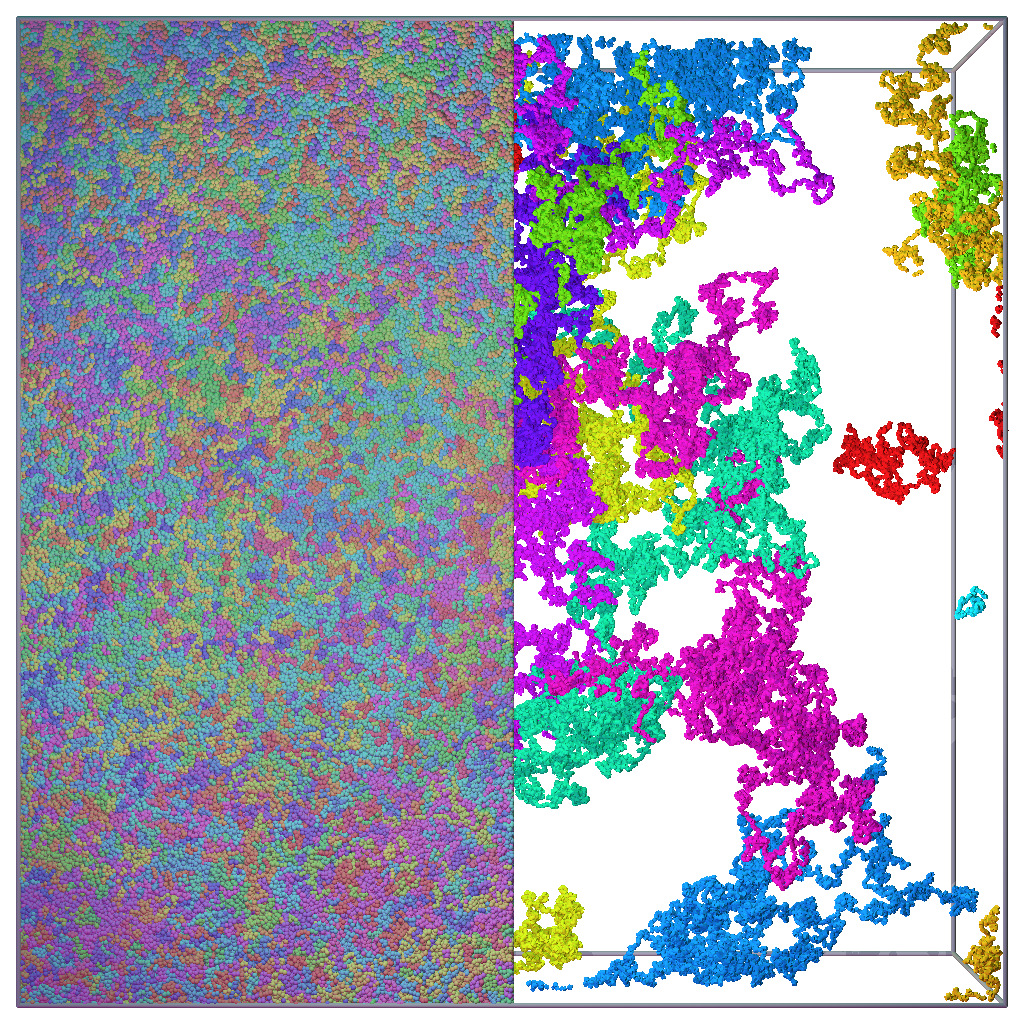}%
\includegraphics[width=0.45\columnwidth]{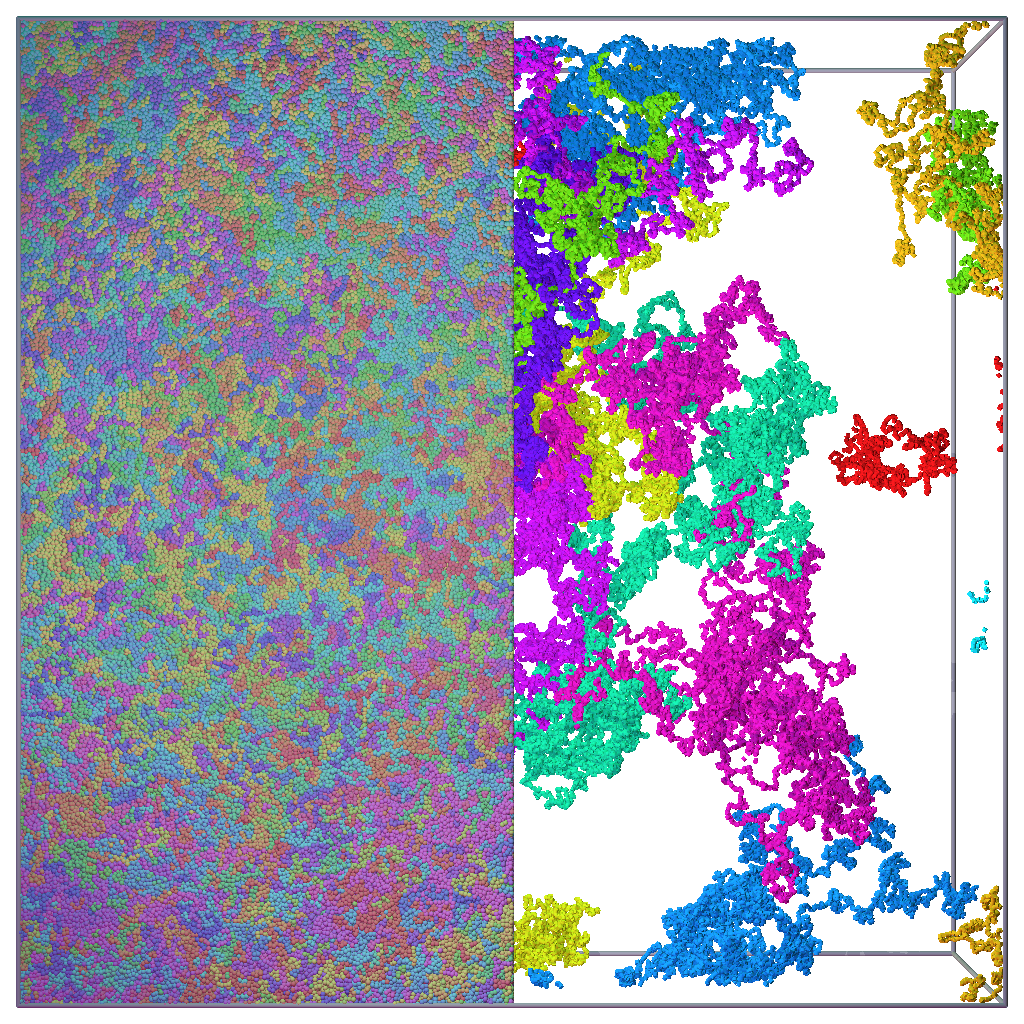}\caption{\label{fig:visualization}Visualization of a $M\times N_{b}=1.000\times15.000$
melt state during the equilibration process. All the polymers are
shown on the left hand side of the box, while the same $10$ randomly
selected polymers are shown on the right hand side of the box. Configuration
just after lattice annealing (top left), after $0.1\tau_{e}$ (top
right), and $1\tau_{e}$ (bottom left) of Rouse dynamics simulation
with the force-capped KG model, and final melt configuration after
KG warm up (bottom right). }
\end{figure}

We borrow ideas from many of the approaches described above, but with
a few twists and improvements. The most important being that we use
different polymer models at the different scales. At the tube length
scale, we model the polymer as a chain of entanglement blobs on a
lattice and minimize spatial density fluctuations using Monte Carlo
simulated annealing. Here we use double-bridge moves, which are easy
to identify and always accepted. The lattice conformations are mapped
onto a bead-spring model. Subsequently we equilibrate the chain structure
at the tube length scale and below using a force capped force field
inspired from dissipative-particle dynamics\citep{hoogerbrugge1992simulating,espanol1995statistical}.
The resulting Rouse dynamics allows chains to pass through each other,
and hence equilibrate local chain structure. Finally we switch to
the Kremer-Grest force field and thermalize the conformations to produce
the correct local bead packing. Each of these steps are fast because
we are using computationally efficient models at each scale. The evolution
of an example configuration is shown in Fig. \ref{fig:visualization}
for a melt with $M=1.000$ chains of length $N_{b}=15.000$ beads. 

The paper is structured as follows; In the short theory section we
introduce the basic concepts and quantities characterizing polymer
melts. In Sect. \ref{sec:Polymer-models}, we define and characterize
the three polymer models that we use in the paper. In Sect. \ref{sec:Characterization-of-equilibratio},
we proceed to characterize the equilibration process in terms of single-chain,
collective, and bulk observables at microscopic, mesoscopic and macroscopic
scales. Finally, we conclude with our conclusion in Sect. \ref{sec:Conclusions}.

\section{Characteristics of polymer melts\label{sec:Theory}}

Below we introduce the characteristic spatial and temporal scales
associated with polymers conformations and their dynamics. At the
molecular scale, we can characterize the single chain statistics in
a polymer melt as a ideal random walk, since excluded volume interactions
are approximately screened.\citep{flory49,flory1989statistical} We
can characterize chain statistics either in terms of number of carbon
atoms in the backbone or number of monomers, however, since our target
here is the Kremer-Grest bead-spring model, we express conformations
in terms of the number of beads $N_{b}$ per chain. The end-to-end
distance of a chain of $N_{b}$ beads is then given by

\begin{equation}
\langle R^{2}(N_{b})\rangle=c_{b}l_{b}^{2}N_{b}=l_{K}L_{K},\label{eq:end-to-end-distance}
\end{equation}
where $l_{b}$ is the average bond length, and $c_{b}$ is the chain
stiffness due bead-packing and local chain structure. For $N_{b}\gg1$
the chain stiffness is given by $c_{b}=\left[\langle\cos\theta\rangle+1\right]/\left[\langle\cos\theta\rangle-1\right]$,
where $\theta$ denotes the angle between subsequent bonds. At the
Kuhn scale (denoted by subscript ``K'') the chain statistics becomes
particular simple. It is described by a random walk with contour length
$L_{K}=l_{K}N_{K}=l_{b}N_{b}$ where the walk consists of $N_{K}$
Kuhn segments that are statistically independent i.e. $c_{K}=1$ at
and above the Kuhn scale. 

The Kuhn length can be estimated using 

\begin{equation}
l_{K}=\frac{\langle R^{2}(N_{b})\rangle}{L_{K}}=2\sqrt{\langle l_{b}^{2}\rangle}\int_{0}^{N_{b}}\left(1-\frac{n}{N_{b}}\right)C(n)\mbox{d}n,\label{eq:Kuhn_estimation}
\end{equation}
where we have expressed the mean-square end-to-end distance in terms
of the bond correlation function $C(n)=\left\langle {\bf b}(m)\cdot{\bf b}(m+n)\right\rangle _{m}$.
This correlation function characterize along how many bonds correlations
between bond directions persists. The bond correlation function is
easy to sample from simulations. 

To define a mesoscopic length scale due to collective chain effects,
we can look at most characteristic macroscopic material property of
a polymer melt -- the plateau modulus. Since polymers can not move
though each other, thermal fluctuations are topologically constrained.
This leads to a localization of the thermal fluctuations inside a
tube-like shape of typical size $d_{T}$.\citep{edwards67} Each topological
entanglement contributes an free energy of $k_{B}T$, and the plateau
modulus is the corresponding free energy density

\[
G_{N}=\frac{4}{5}\frac{\rho_{K}kT}{N_{eK}}.
\]

Here $\rho_{K}=\rho_{b}/c_{b}$ is the number density of Kuhn segments,
$\rho_{b}$ the number density of beads, $k$ is the Boltzmann constant,
and $T$ the temperature. The entanglement length $N_{ek}$ is a measure
of the contour length between topological entanglements along the
chain. Note that we specify it in terms of Kuhn units and not beads
between entanglements. In the present paper, we generally report results
in terms of Kuhn units rather than numbers specific for the Kremer-Grest
model. This is to simplify comparisons with theory and experiment,
since in Kuhn units we would characterize a real chemical molecule
and one of our model molecules with exactly the same numbers independent
of the chosen polymer model. The $4/5$ pre-factor is due to the entanglements
lost as the stretched chains initially retract into the tube to reestablish
their equilibrium contour length.\citep{DoiEdwards_86}

We can relate the length of a tube segment $d_{T}$ to the number
of Kuhn units it contains as $d_{T}^{2}=l_{K}^{2}N_{eK}$ and $Z=N_{K}/N_{eK}$
as the number of entanglements/tube segments per chain. Since the
tube is a coarse representation of the chain it contains, the large
scale tube and chain statistics must coincide, while below the tube
length scale, the tube is straight and the chain performs a random-walk.
In particular, the chain end-to-end distance matches the end-to-end
distance of the tube $\langle R^{2}\rangle=d_{T}^{2}Z=l_{K}^{2}N_{K}$.

The dynamics of short unentangled polymer melts, is described by the
Rouse model,\citep{rouse1953theory,DoiEdwards_86} which also describes
the local dynamics of long entangled melts. In this model, a chain
is represented by a flexible string of non-interacting units connected
by harmonic springs, i.e. each unit represents one Kuhn segment of
the polymer. Besides the forces that arise due to connectivity, each
unit also receives a stochastic kick and is affected by a friction
force, i.e. the Rouse model is endowed with Langevin dynamics. The
combined effects of these two forces are to model the presence of
the other chains in the melt. The Rouse model can be solved exactly
analytically by transforming it to a mode representation, see e.g.
\citep{DoiEdwards_86}. In particular, the Rouse model predicts the
chain centre-of-mass diffusion coefficient $D_{cm}$ and its relation
to the Kuhn friction $\zeta_{K}$ as

\begin{equation}
D_{cm}=\frac{kT}{\zeta_{K}N_{K}}.\label{eq:dcm}
\end{equation}
which has the form of a fluctuation-dissipation theorem. This relation
can be inverted to derive the Kuhn friction from a measured diffusion
coefficient. The fastest dynamics is that associated with the diffusive
motion of individual Kuhn segments one Kuhn length i.e. $\tau_{K}\sim l_{K}^{2}D_{K}^{-1}\sim\zeta_{K}l_{K}^{2}/kT$.
A more careful derivation using Rouse theory provides the prefactor
as

\begin{equation}
\tau_{K}=\frac{\zeta_{K}l_{K}^{2}}{3\pi^{2}kT}.\label{eq:tau_k}
\end{equation}

In the case of entangled melts, we can define the entanglement time
which is the characteristic time it takes an entangled chain segment
to diffuse the length of a tube segment $\tau_{e}\sim d_{T}^{2}(D_{K}/N_{e})^{-1}\sim l_{K}^{2}N_{e}^{2}\zeta_{K}/kT$,
and with prefactors from Rouse theory

\begin{equation}
\tau_{e}=\tau_{K}N_{e}^{2}=\frac{\zeta_{K}}{3\pi^{2}kT}\frac{d_{T}^{4}}{l_{K}^{2}},\label{eq:taue}
\end{equation}
the entanglement time is typically much larger than the fundamental
Kuhn time. The conformational relaxation times due to reptation (linear
polymers) or contour length fluctuations (star polymers) is again
typically much larger than the entanglement time. 

The Kuhn length is a microscopic single chain property, and the tube
diameter is a collective mesoscale property that is typically associated
with pair-wise entanglements.\citep{everaers2012topological} In order
to characterize bulk large scale melt properties and in particular
density fluctuations, we use the structure factor. The structure factor
is defined as
\begin{equation}
S({\bf q})=\left(N_{b}M\right)^{-1}\left\langle \left|\sum_{j=1}^{M}\sum_{k=1}^{N_{b}}\exp(i{\bf q}\cdot{\bf R}_{jk})\right|^{2}\right\rangle ,\label{eq:structurefactorpoint}
\end{equation}
where ${\bf q}$ is the momentum transfer in the scattering process.
$M$ denotes the number of polymers, and ${\bf R}_{jk}$ the position
of the $k$'th bead in the $j$'th polymer. We assume for notational
simplicity that all polymers has the same number of beads. When performing
simulations with periodic boundary conditions, we are limited to momentum
transfers on the reciprocal lattice of the simulation box i.e. ${\bf q}$
vectors of the form ${\bf q}=(2\pi n_{x}/L,2\pi n_{y}/L,2\pi n_{z}/L)$,
where $L$ denote the box size. Since the melts are isotropic, we
average and bin the structure factor based on the magnitude of the
momentum transfer vector denoted $q=|{\bf q}|$. The structure factor
for small $q$ values converges to $\lim_{q\rightarrow0}S(q)=\chi_{T}\rho kT$
where $\chi_{T}$ is the isothermal compressibility of the melt. For
a further discussion on density fluctuations and compressibility,
we refer to the more detailed derivations in the appendix.

\section{Polymer models\label{sec:Polymer-models}}

In the following, we define and characterize the three polymer models
employed in the present study: We begin with the Kremer-Grest model
(sec. \ref{sub:Kremer-Grest-polymer-model}), we also introduce a
force capped variant of the Kremer-Grest model (fcKG) (sec. \ref{sub:Auxiliary-polymer-model}),
and finally we introduce a model where chains are modelled as a string
of entanglement blobs on a lattice (sec. \ref{sub:Lattice-blob-model}).
We also characterize the Kuhn length for both the KG and fcKG models
(sec. \ref{sub:Kuhn-length}), the tube diameter for the Kremer-Grest
model (sec. \ref{sub:Tube-diameter-of}), and finally the Kuhn friction
of the fcKG model (sec. \ref{sub:Time-mapping-of}). These relations
are required to transfer melt conformations between the different
polymer models, and to determine how long a Rouse simulation is required
for the equilibration process.

\subsection{Kremer-Grest polymer model\label{sub:Kremer-Grest-polymer-model}}

The end goal of the present equilibration procedure is to produce
an equilibrated Kremer-Grest model melt.\citep{grest1986molecular,kremer1990dynamics}
This is a generic bead-spring polymer model, where all beads interact
via a Weeks-Chandler-Anderson (WCA) potential

\[
U_{WCA}=4\epsilon\left[\left(\frac{\sigma}{r}\right)^{-12}-\left(\frac{\sigma}{r}\right)^{-6}+\frac{1}{4}\right]\quad\mbox{for}\quad r<2^{1/6}\sigma,
\]
 while springs are modeled by finite-elastic-non-extensible spring
(FENE) potential

\[
U_{FENE}=-\frac{kR^{2}}{2}\ln\left[1-\left(\frac{r}{R}\right)^{2}\right],
\]
where we choose $\epsilon$ and $\sigma$ as the units of energy and
distance respectively. The unit of time is $\tau=\sigma\sqrt{m_{b}/\epsilon}$
where $m_{b}$ denotes the mass of a bead. We add an additional bending
interaction given by

\[
U_{bend}(\Theta)=\kappa\left(1-\cos\Theta\right).
\]

The bending potential was introduced by Faller and Müller-Plathe.\citep{faller1999local,faller2000local,faller2001chain}
The KG models are simulated using Langevin dynamics, which couples
all beads to a thermostat, and allows long simulations at constant
temperature to be performed with reasonable large time steps. The
Langevin dynamics is given by the conservative force due pair and
bond interactions, as well as a friction term and a stochastic force
term:

\[
m\frac{\partial^{2}{\bf R}{}_{n}}{\partial t^{2}}=-\nabla_{{\bf R}_{n}}U-\Gamma\frac{\partial}{\partial t}{\bf R}_{n}+{\bf \xi}_{n},
\]
where the stochastic force obeys $\langle{\bf \xi}_{n}\rangle=0$
and $\langle{\bf \xi}_{n}(t)\cdot{\bf \xi}_{m}(t')\rangle=6kT\Gamma\delta(t-t')\delta_{nm}$.
The standard choice of the FENE bonds are $R=1.5\sigma$ and $k=30\epsilon\sigma^{-2}$,
which produce a bond length of $l_{b}=0.965\sigma$\citep{subramanian2010topology}.
Hence the number of beads per Kuhn unit is $c_{b}=l_{K}(\kappa)/l_{b}$.
The standard value for the thermostat coupling is $\Gamma=0.5m_{b}\tau^{-1}$.
KG model melts are typically simulated with a bead density of $\rho_{b}=0.85\sigma^{-3}$.
We use a time step of $\Delta t=0.01\tau$. For integrating the dynamics
of our force field, we utilize the Grønbech-Jensen/Farago Langevin
integration algorithm\citep{gronbech2013simple,gronbech2014application}
implemented in the Large Atomic Molecular Massively Parallel Simulator
(LAMMPS).\citep{PlimptonLAMMPS}

\subsection{Force-capped KG model\label{sub:Auxiliary-polymer-model}}

The KG model preserves topological entanglements via a kinetic barrier
of about $75kT$ for chain pairs to move through each other.\citep{sukumaran2005identifying}
This is due to the strong repulsive pair interaction in combination
with a strongly attractive bond potential that diverges when bonds
are stretched towards the maximal distance $R$. Preserving topological
entanglements is essential for reproducing the plateau modulus. The
lattice melt configurations has the correct large scale chain statistics,
but as we will show later, the density of entanglements is much too
low, hence directly switching from a lattice configuration to a topology
preserving KG polymer model would produce model melts with a wrong
entanglement density. Hence we need a computationally effective model
to introduce the correct random walk statistics inside the tube diameter,
and hence produce the correct entanglement density before switching
to the KG model.

The force-capped Kremer-Grest model (fcKG) should solve this problem
by 1) performing a Rouse like dynamics to introduce local random walk
chain statistics, 2) prevent the growth of density fluctuations, 3)
avoid the numerical instabilities due to short pair distances or long
bonds which can occur in the lattice melt state or during the Rouse
dynamics of the fcKG  model, and finally 4) approximate the ground
state of the KG force field such we can switch to this force field
with a minimum of computational effort.

Inspired from dissipative particle dynamics\citep{hoogerbrugge1992simulating,espanol1995statistical}
and a previous equilibration methods \citep{zhang2014equilibration,moreira2015direct},
we apply a force-cap to the WCA potential as follows

\begin{equation}
U_{WCA}^{cap}(r)=\begin{cases}
(r-r_{c})\left.\frac{dU_{WCA}}{dr}\right|_{r=r_{c}}+U_{WCA}(r_{c}) & \quad\quad r<r_{c}\\
U_{WCA}(r) & \quad\quad r_{c}\leq r\leq2^{1/6}\sigma\\
0 & \quad\quad r>2^{1/6}\sigma
\end{cases}.\label{eq:forcecap}
\end{equation}

The inner cutoff distance $r_{c}$ determines the potential at overlap.
We choose $U_{WCA}^{cap}(r=0)=5\epsilon$ which corresponds to an
inner cutoff of $r_{c}=0.9558\times2^{1/6}\sigma$. For the bond potential,
we choose a fourth degree Taylor expansion of sum of the original
WCA and FENE bond potentials around the equilibrium distance ($r_{0}=0.9609\sigma$).
The resulting bond potential is

\[
U_{bond}(r)=20.2026\epsilon+490.628\epsilon\sigma^{-2}(r-r_{0})^{2}
\]

\begin{equation}
-2256.76\epsilon\sigma^{-3}(r-r_{0})^{3}+9685.31\epsilon\sigma^{-4}(r-r_{0})^{4}.\label{eq:feneapprox}
\end{equation}

Finally we retain the bending potential

\[
U_{bend}(\Theta)=\kappa_{fc}\left(1-\cos\Theta\right),
\]
and simulate the fcKG model with exactly the same Langevin dynamics
as the full KG model. 

We avoid numerical instabilities by using the Taylor expansion in
the fcKG model rather than FENE and WCA potentials between bonded
beads in the KG model. As a result the numerical stability of the
force capped model is considerably improved both for very short and
very long bonds. We can simulate the lattice melt states directly
(after simple energy minimization) without requiring any elaborate
push-off or warm-up procedures to gradually change the force field.
Since the force-capped model also approximates the ground state of
the full KG model, we can also switch force-capped melt configurations
to the full KG force field using simple energy minimization and also
avoid designing a delicate push-off or warm-up procedure for this
change of force field. Furthermore, expect an increased bead mobility
while local single chain structure remains mostly unaffected. Note
that in the KG model the WCA interaction is applied between all bead
pairs, however for the fcKG model the WCA potential is already included
in Taylor expanded bond potential above, hence the pair interaction
is limited to non-bonded beads for the force capped KG model. 

\begin{figure}
\includegraphics[width=0.33\columnwidth]{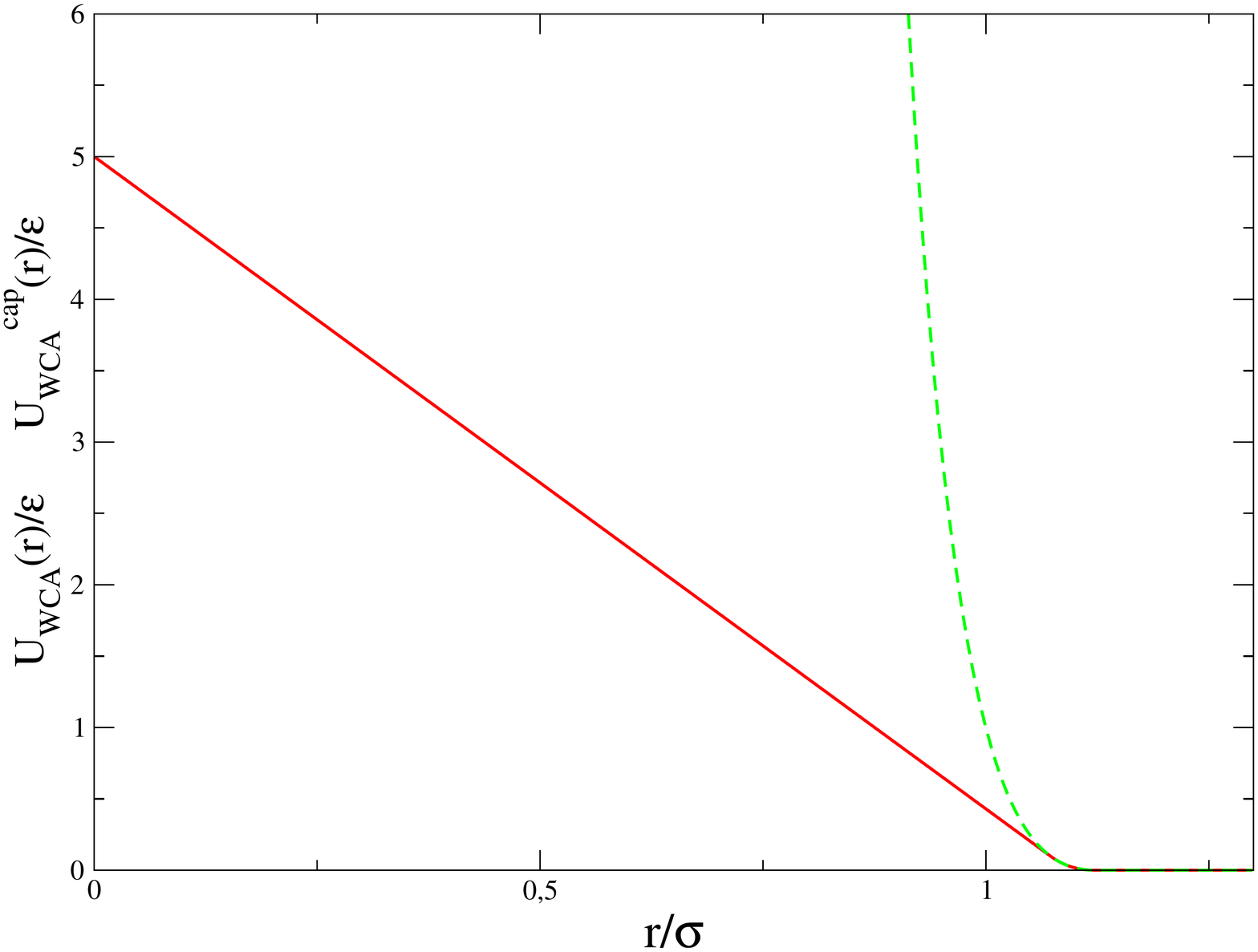}%
\includegraphics[width=0.33\columnwidth]{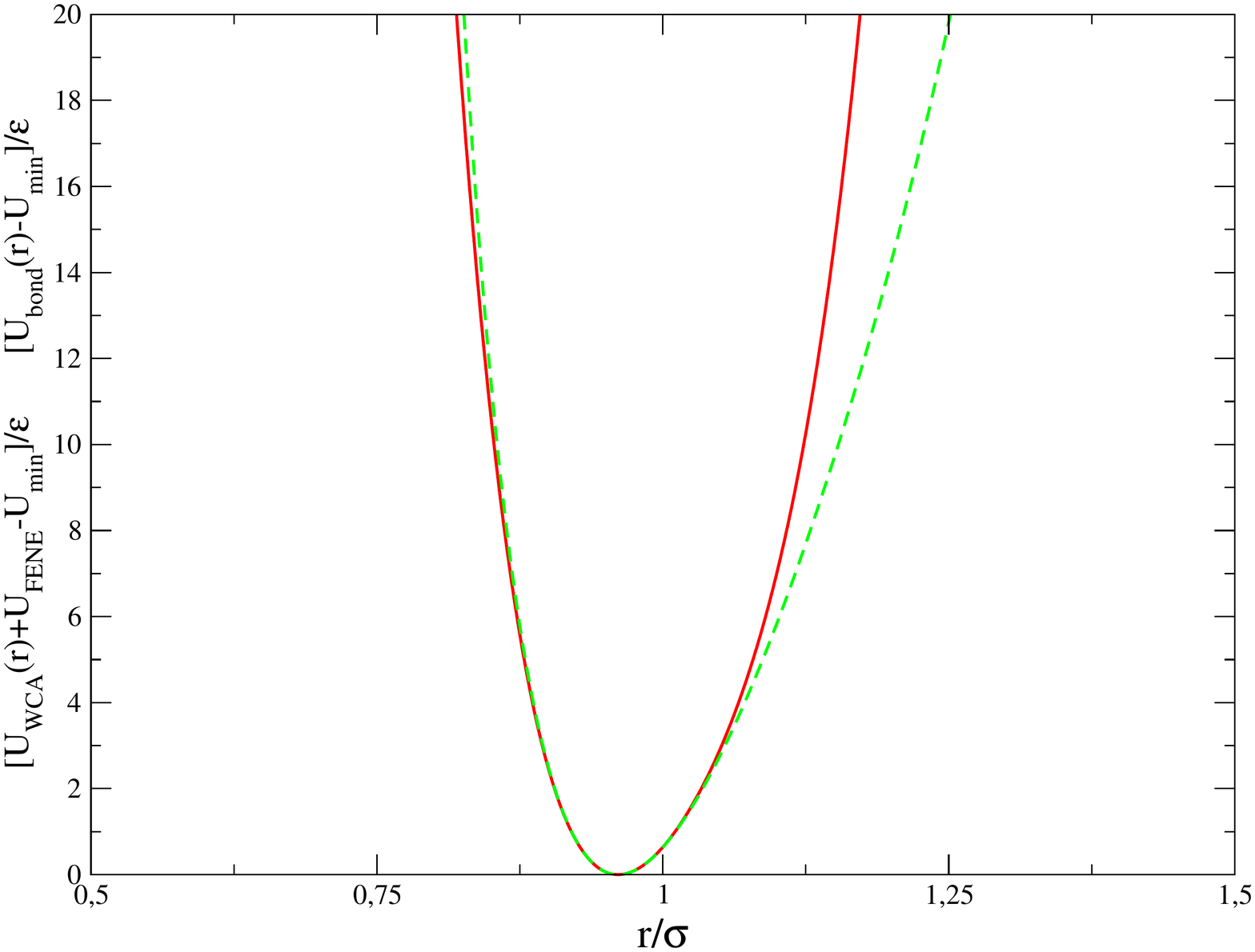}%
\includegraphics[width=0.33\columnwidth]{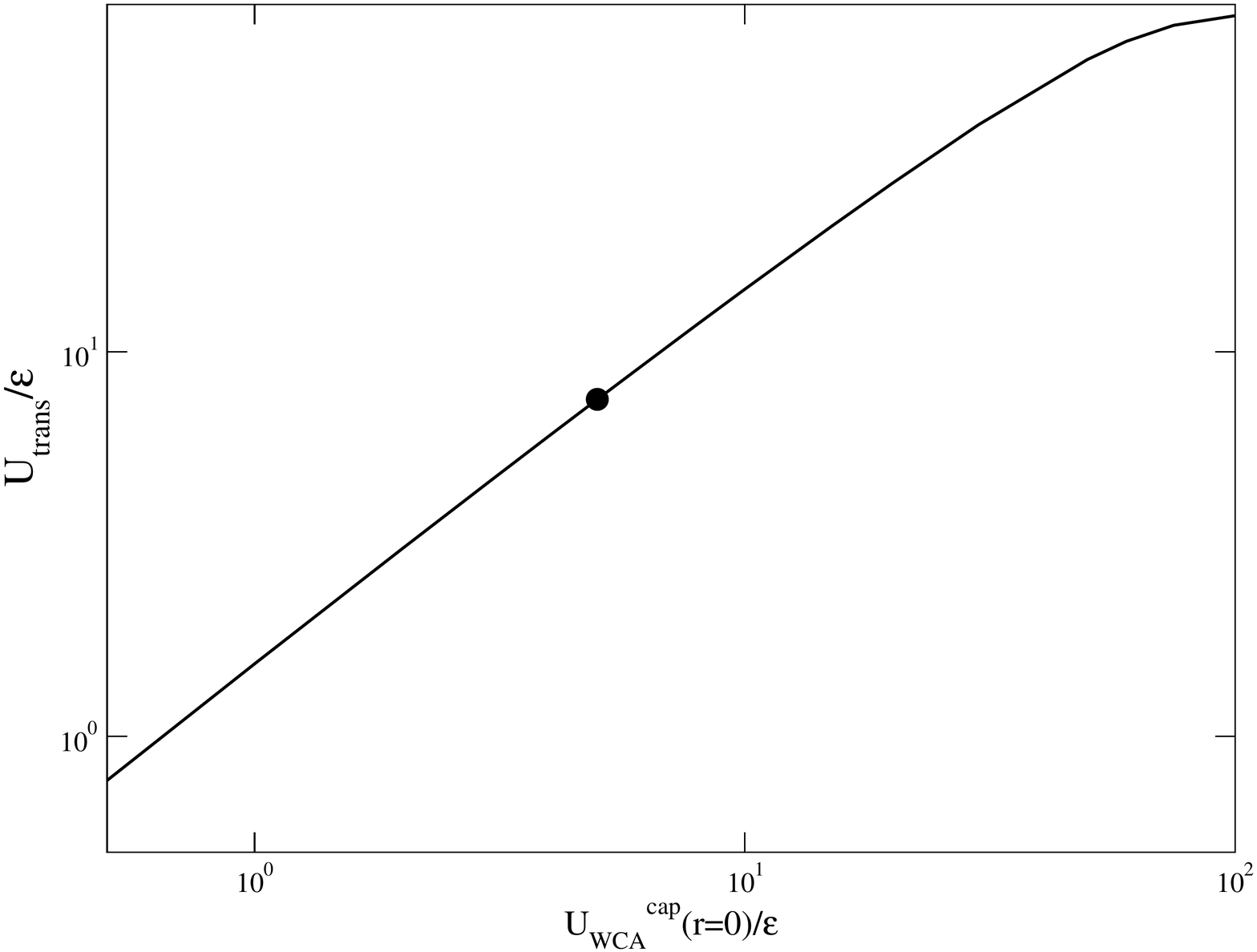}
\caption{The pair potential (left) and bond potential (middle) for the full
KG model (green dashed lines) and for the fcKG model (red solid lines).
Shown is also the height of the energetic barrier for chains to pass
through each other as function of the force cap (right). The circle
denotes our choice of $U_{WCA}^{cap}(r=0)=5\epsilon$\label{fig:Potentials}}
\end{figure}

Figure \ref{fig:Potentials} shows a comparison between the pair and
bonded potentials of the KG and fcKG models. The figure also shows
the height of the energy barrier of chains passing through each other
as function of the force cap expressed as function of the pair potential
at overlap $U_{WCA}^{cap}(r=0)$. The transition state is a planar
configuration of two perpendicular chains, where two perpendicular
bonds open up to allow one chain to pass through the other. Compared
to the KG model, this force cap reduce this energy barrier from $75\epsilon$
down to $7.5\epsilon$. 

\subsection{Lattice blob model\label{sub:Lattice-blob-model}}

We coarse-grain space into a lattice on a length scale $a$ corresponding
to the tube segment length $d_{T}$. The polymers become random walks
on this lattice. Since multiple chains pervade an entanglement volume,
multiple blobs can occupy the same lattice site. We regard the polymers
as consisting of $Z$ entanglement blobs of $N_{e}$ Kuhn segments
each. The number of chains within the volume associated with a blob
is $n_{e}=\rho_{K}N_{e}^{-1}d_{T}^{3}$. For most flexible well-entangled
polymers $n_{e}\sim19$.\citep{fetters2007chain} 

We utilize the recently published lattice polymer model of Wang\citep{wang2009studying}.
This model has the computational advantage that the Hamiltonian does
not include pair-interactions, which makes it computationally very
effective. The Hamiltonian comprises an incompressibility term and
a configuration term as follows

\begin{equation}
H=\frac{1}{2\chi\langle n\rangle}\sum_{c}\left(n_{c}-\langle n\rangle\right)^{2}+\sum_{p}\left(\epsilon_{0}N_{p0}+\epsilon_{90}N_{p90}+\epsilon_{180}N_{p180}\right),\label{eq:Hamiltonian}
\end{equation}
the first term is a sum over all sites, while the second is a sum
over all polymers. $n_{c}$ denotes the blob occupation number at
site $c$, while $\langle n\rangle\approx n_{e}$ is the average number
of blobs per site. The parameter $\chi$ plays the role of a compressibility\citep{helfand1971theory,helfand1972theory}
and hence allows us to introduce incompressibility gradually to remove
large scale density fluctuations. In the configuration term we sum
over bond angles in the chains. The three terms represents anti-parallel,
orthogonal, and parallel successive bonds and their respective energy
penalties, respectively. The average bond-bond angle is in this case
given by

\[
\langle\cos\Theta\rangle=\frac{-\exp(-\beta\epsilon_{0})+\exp(-\beta\epsilon_{180})}{\exp(-\beta\epsilon_{0})+4\exp(-\beta\epsilon_{90})+\exp(-\beta\epsilon_{180})},
\]
to obtain non-reversible random walk of blobs we require $\langle\cos\Theta\rangle=0$,
such that $c_{L}=\left[\langle\cos\theta\rangle+1\right]/\left[\langle\cos\theta\rangle-1\right]=1$.
We choose the parameters $\epsilon_{0}=\epsilon_{180}=1$ and $\epsilon_{90}=0$.
We furthermore choose $\chi=1$. Since we are doing simulated annealing
the exact values of these parameters are irrelevant. Any state with
density fluctuations or configurations with deviations from non-reversible
random walks will be exponentially unlikely when the temperature is
reduced sufficiently. 

We have implemented double bridging, pivot, reptation, and translate
moves. Double bridge moves are performed by identifying two pairs
of connected blobs on neighboring sites where ``crossing over''
the bond between the two pairs of blobs does not change mono-dispersity
of the melt. Since double bridge moves alters neither angles nor blob
positions, the double bridge moves does not change the energy, and
are always accepted. Double bridge moves can be carried out both inside
a chain and between pairs of chains. Pivot moves picks a random bond
and randomly pivots the head or tail of the chain around the the chosen
bond.\citep{madras1988pivot} Pivot moves only change one angle at
the pivot point, but cause major spatial reorganization of the polymer.
In densely packed systems, the acceptance rate of pivot moves drops
rapidly. Reptation moves delete an number of blobs at either the head
or the tail of a polymer and regrows the same number of blobs at the
other end of the polymer. Reptation moves are very efficient at generating
new configurations in dense systems. Translate moves pick a random
bond and randomizes it, and hence randomly translates the head or
the tail of the chain by one lattice step relative to the bond. Of
the moves discussed here, only the reptation move is limited to linear
chain connectivity. We implemented the Metropolis Monte Carlo algorithm
in C++ (2011 standard version) making extensive use of standard-template
library containers and pointer structures choosing optimal data structures
for implementing the infrastructure for generating new moves, rejecting
moves with a minimal overhead, and rapidly estimating the energy change
of a given trial move.

We note that our choice of lattice length scale $a$ is in fact an
arbitrary, since the subsequent Rouse simulation with the fcKG model
removes the lattice artifacts. From eq. \ref{eq:taue}, we see that
the Rouse simulations duration growths as the fourth power of the
lattice constant. On the other hand, the advantage of enforcing the
incompressibility constraint with a lattice Hamiltonian requires a
meaningful site occupation numbers $n_{c}\gg1$. When this limit is
approached, the incompressibility constraint converges to an excluded
volume constraint and blobs to single monomers. Matching the lattice
spacing and the tube diameter producing $\langle n\rangle\sim19$
offers a reasonable compromise.

\subsection{Kuhn lengths of both KG models\label{sub:Kuhn-length}}

In order to have the same chain statistics and in particular a specific
Kuhn length for the force capped and full KG models, we need to estimate
how these change with stiffness. Theoretically predicting the Kuhn
length of a polymer model with pair-interactions is a highly non-trivial
problem. While excluded volume interactions are approximately screened
in melts (the Flory ideality hypothesis\citep{flory49,flory1989statistical}),
melt deviates from polymers in $\Theta$-solutions due to their incompressibility.
The incompressibility constraint creates a correlation hole, which
leads to a long range net repulsive interaction between polymer blobs
along the chain, this effectively causes a renormalization of the
bead-bead stiffness to make them stiffer.\citep{wittmer2007polymer,wittmer2007intramolecular,beckrich2007intramolecular,meyer2008static,semenov2010bond}

To circumvent this problem, we have brute force equilibrated medium
length entangled melts with $M=2000$ chains of length $N_{b}=400$
beads while systematically varying the stiffness parameter for both
the KG and fcKG models. Each initial melt conformation was simulated
for at least $2\times10^{5}\tau$ while performing double-bridging
hybrid MC/MD simulations\citep{daoulas2003variable,karayiannis2002atomistic,karayiannis2002novel,karayiannis2003advanced}
using the bond/swap fix in LAMMPS\citep{sides2004effect}. Ten to
twenty configurations from the last $5\times10^{4}\tau$ of the trajectory
were used to estimate the Kuhn length. We choose the chain length
as a compromise between having as many Kuhn segments as possible and
on having an acceptable double bridging acceptance rate. While double
bridging moves are very efficient at removing correlations between
the chain conformations, the acceptance rate drops significantly with
chain lengths since the potential cross-over points are progressively
diluted when requiring that the melt remains mono-disperse. The Kuhn
length were derived using eq. \ref{eq:Kuhn_estimation}. 

\begin{figure}
\includegraphics[width=0.95\columnwidth]{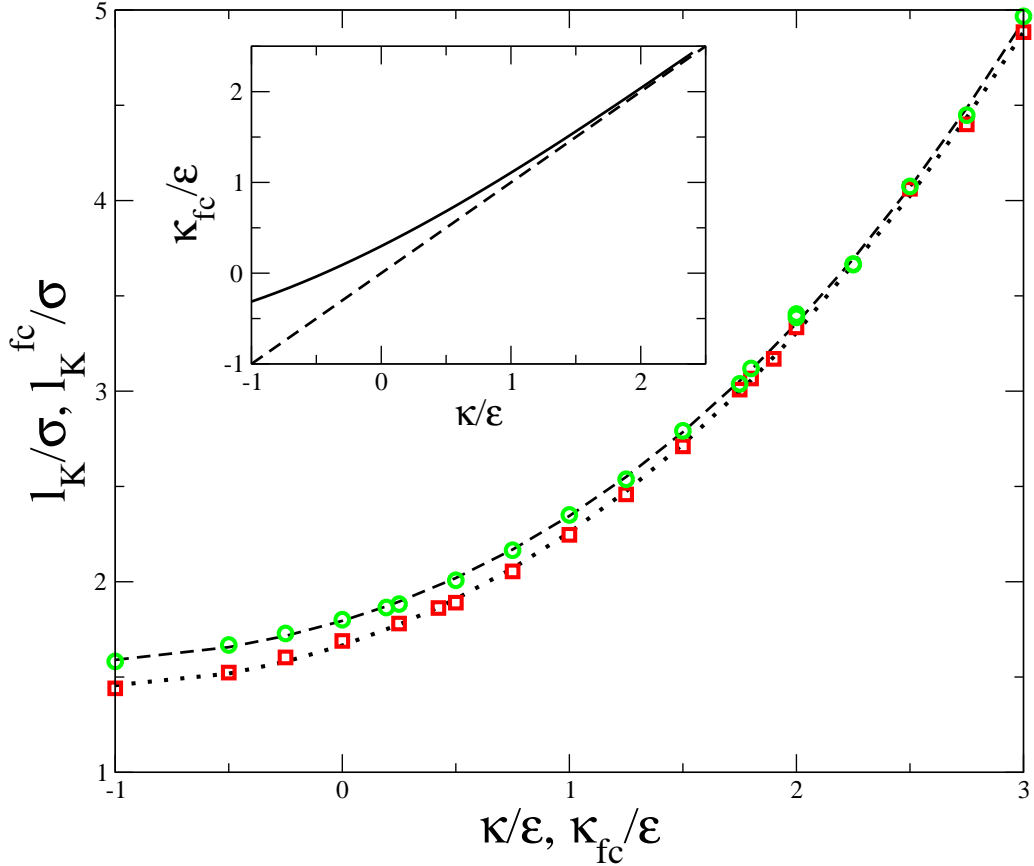}

\caption{\label{fig:Kuhn-length}Kuhn length $l_{K}$ vs stiffness parameter
for the KG (green circles) and fcKG models (red boxes). The lines
are polynomial fits $l_{K}(\kappa)/\sigma=1.795+0.358\epsilon^{-1}\kappa+0.172\epsilon^{-2}\kappa^{2}+0.019\epsilon^{-3}\kappa^{3}$
(hashed black line) and $l_{K}^{fc}(\kappa_{fc})/\sigma=1.666+0.389\epsilon^{-1}\kappa_{fc}+0.192\epsilon^{-2}\kappa_{fc}^{2}+0.012\epsilon^{-3}\kappa_{fc}^{3}$
(dotted black line). The insert shows the relation between $\kappa_{fc}$
and $\kappa$ defined by eq. \ref{eq:kapparenorm}. (solid black line)}
\end{figure}

The resulting Kuhn lengths are shown in Fig. \ref{fig:Kuhn-length}.
As expected, as the stiffness parameter is increased the Kuhn length
grows concomitantly. The stiffness of the fcKG and the KG models varies
slightly. This is due to the additional stiffness introduced by the
WCA pair interaction between next nearest neighbors along the chain
compared to the force-capped model. Using the extrapolations shown
in Fig \ref{fig:Kuhn-length} we can numerically solve for the force-capped
model stiffness $\kappa_{fc}$ required to reproduce equivalent KG
model with stiffness parameter $\kappa$. The result is shown in Fig.
\ref{fig:Kuhn-length}, and is given by the following empirical relationship
valid for $\kappa\in[-1:2.5]$. 

\begin{equation}
\kappa_{fc}(\kappa)=0.298+0.722\kappa+0.099\kappa^{2}-0.012\kappa^{3}.\label{eq:kapparenorm}
\end{equation}

\subsection{Tube diameter of Kremer-Grest melts\label{sub:Tube-diameter-of}}

In order to choose the spacing of the lattice model, we need to estimate
the length of a tube segment $a(\kappa)$ as function of stiffness
$\kappa$ for the KG model. We have generated $15$ melt states with
$M=500$ chains of length $N_{b}=10.000$ for $\kappa=-1,-0.75,-0.50,\cdots,2.25,2.50$.
We used the algorithm of the present paper, but chose the lattice
spacing $a=l_{K}(\kappa)\sqrt{N_{K}(\kappa)}$, with $N_{K}(\kappa)=100c_{b}^{-1}(\kappa)$.
This corresponds to using not entanglement blobs, but rather blobs
with a fixed number of beads ($100$) independently of chain stiffness.

We have performed primitive-path analysis (PPA) of the melt states.\citep{everaers2004rheology}
During the PPA a melt conformation is converted into the topologically
equivalent primitive-path mesh-work characterizing the tube structure.
We have performed a version of the PPA analysis which preserves self-entanglements
by only disabling pair interactions between beads within a chemical
distance of $2c_{b}N_{eK}$ bonds.\citep{sukumaran2005identifying}
The minimization was performed using the steepest descent algorithm
implemented in LAMMPS followed by dampened Langevin dynamics as described
in Ref. \citep{everaers2004rheology}.

\begin{figure}
\includegraphics[width=0.9\textwidth]{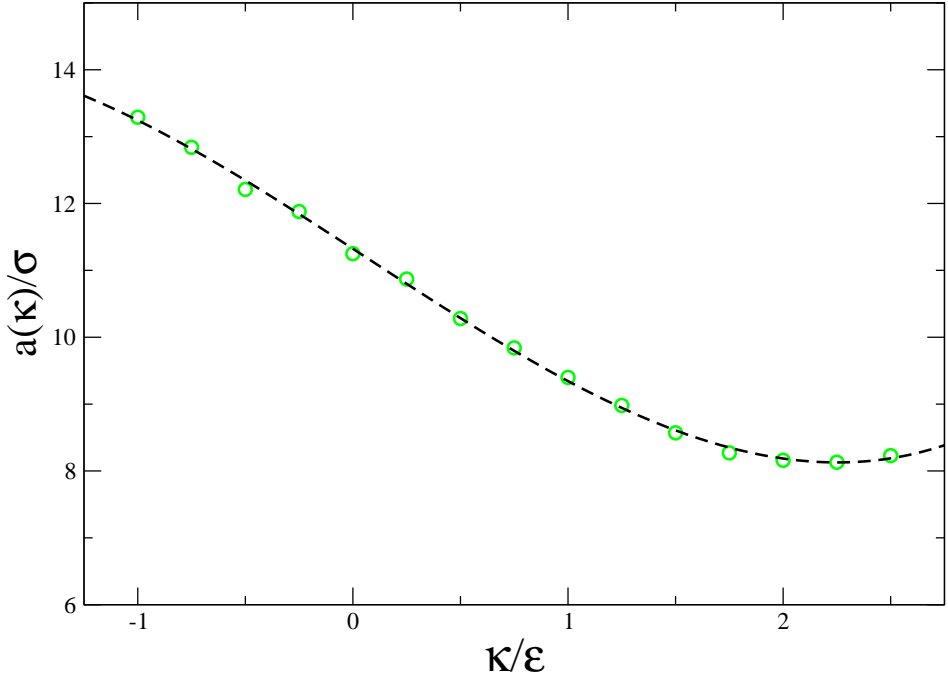}

\caption{\label{fig:tubediameter}Tube segmental length $a(\kappa)$ vs stiffness
parameter for the KG model (green symbols), also shown is an interpolation
given by $a(\kappa)/\sigma=11.32-2.096\epsilon^{-1}\kappa-0.0293\epsilon^{-2}\kappa^{2}+0.1465\epsilon^{-3}\kappa^{3}$}
\end{figure}

Since the large scale chain melt statistics and primitive-path statistics
agree, the PPA essentially consists of filtering out the effects of
thermal fluctuations on the chain configurations. The chain mean-square
end-to-end distance is constant, and hence the Kuhn length of the
tube (the tube diameter) is given by $a=\langle R^{2}\rangle/L_{pp}$.
$L_{pp}$ is the average primitive-path contour length, which we obtain
directly from the mesh work produced by the PPA. By preforming the
analysis on melts of varying $\kappa$ we can obtain the tube diameter
as function of chain stiffness $a(\kappa)$. The result is shown in
Fig. \ref{fig:tubediameter}, and as expected, when the chains becomes
stiffer they can pervade a large volume and hence become more entangled,
which corresponds to the observed decrease of the tube diameter. The
generated melts range from $Z(\kappa=-1)=80$ to $Z(\kappa=2.5)=540$
entanglements. Hence, they are all deep inside the regime of strongly
entangled polymer melts, where we can neglect the effect of ends during
the PPA process.\citep{HoyPRE09}

\subsection{Time mapping of the force-capped KG model\label{sub:Time-mapping-of}}

In order to estimate how long time we should run the Rouse simulation
to relax chain statistics up to the tube scale, we need to know the
entanglement time of the fcKG model. The unit of time of the simulated
force field is $\tau$, however, this unit has no direct relation
to the time scales characterizing the emergent polymer dynamics, which
depends on the force field as well as the thermostat parameters. To
define a natural time scale for the polymer dynamics, we obtain the
effective Kuhn friction $\text{\ensuremath{\zeta}}_{K}$. We have
measured the center-of-mass (CM) diffusion coefficient by performing
a series of simulations with varying stiffness parameter $\kappa_{fc}$.
Each melt contains $2.000$ molecules and chain length $N_{K}=10,20,30,40$.
The melts were equilibrated for a period of $10^{4}\tau$ using double
bridging hybrid MC/MD\citep{daoulas2003variable,karayiannis2002atomistic,karayiannis2002novel,karayiannis2003advanced}.
The resulting equilibrium states were run for up to $2-10\times10^{5}\tau$
and the center of mass diffusion coefficient $D_{cm}(\kappa_{fc},N_{K})$
was obtained from the plateau of the measured mean-square displacements
$D_{cm}(\kappa_{fc},N_{K};t)=\langle[R_{cm}(t)-R_{cm}(0)]^{2}\rangle/[6t]$
for $t>10^{5}\tau$ by sampling plateau values for log-equidistant
times, and discarding simulations where the standard deviation of
the samples exceeded $2\%$ of their average value. 

\begin{figure}
\includegraphics[width=0.9\columnwidth]{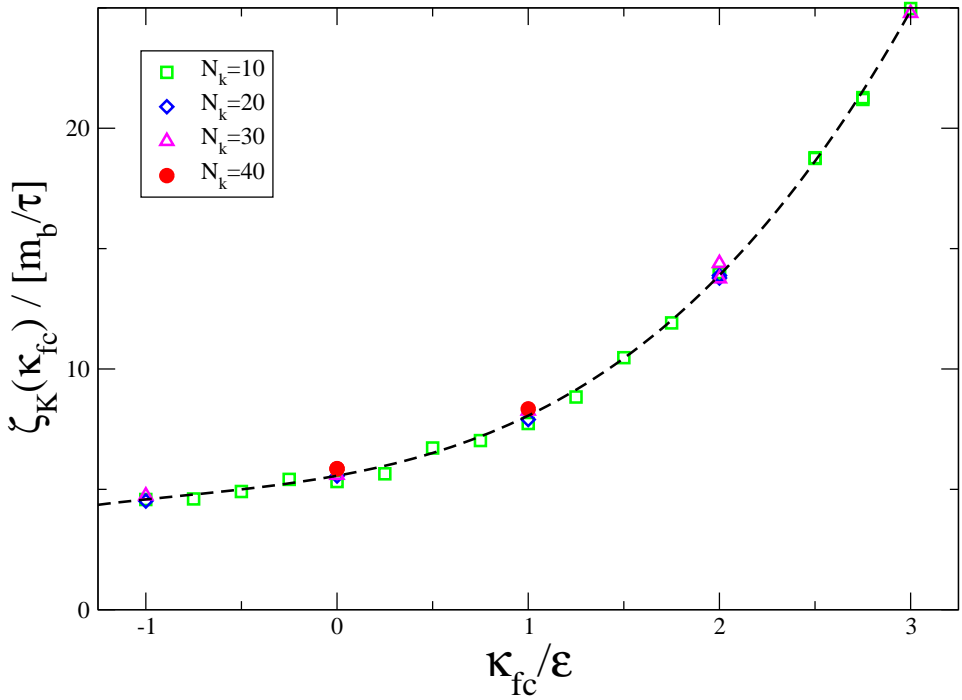}

\caption{\label{fig:Kuhn-friction}Kuhn friction for the fcKG model as function
of stiffness parameter $\kappa_{fc}$. The line through the data points
is the fit $\zeta(\kappa_{fc})/[m_{b}/\tau]=5.5657+1.4367\epsilon^{-1}\kappa_{fc}+0.7564\epsilon^{-2}\kappa_{fc}^{2}+0.30372\epsilon^{-3}\kappa_{fc}^{3}$.}
\end{figure}

Figure \ref{fig:Kuhn-friction} shows the Kuhn friction obtained from
the analysis of the simulations using eq. \ref{eq:dcm}. We observe
that the friction increases slowly with chain stiffness. The excellent
collapse of data from different chain lengths supports the validity
of the Rouse dynamics for the force-capped KG model.

Using Eq. \ref{eq:taue} and the empirical relations shown in Figs.
\ref{fig:tubediameter}, \ref{fig:Kuhn-length}, and \ref{fig:Kuhn-friction},
we obtain an empirical relation for the entanglement time of the fcKG
model as

\begin{equation}
\tau_{e}(\kappa_{fc})/\tau=935.5-710.8\epsilon^{-1}\kappa_{fc}+226.6\epsilon^{-1}\kappa_{fc}^{2}-26.61\epsilon^{-1}\kappa_{fc}^{3},\label{eq:entanglement-time}
\end{equation}
valid within the range of $\kappa_{fc}=-1,\dots,2.5$. The entanglement
time varies from $1900\tau$ down to $160\tau$ as chains gets stiffer.
For simplicity, we define $\tau_{e}(\kappa)\equiv\tau_{e}(\kappa_{fc}(\kappa))$
below to simplify the notation. This does not lead to confusion since
in the present paper entanglement times always refers to the Rouse
simulations of the force capped model, and the aim of running a fcKG
model to equilibrate the corresponding KG models with a specific stiffness
$\kappa$.

\begin{figure}
\includegraphics[width=0.9\columnwidth]{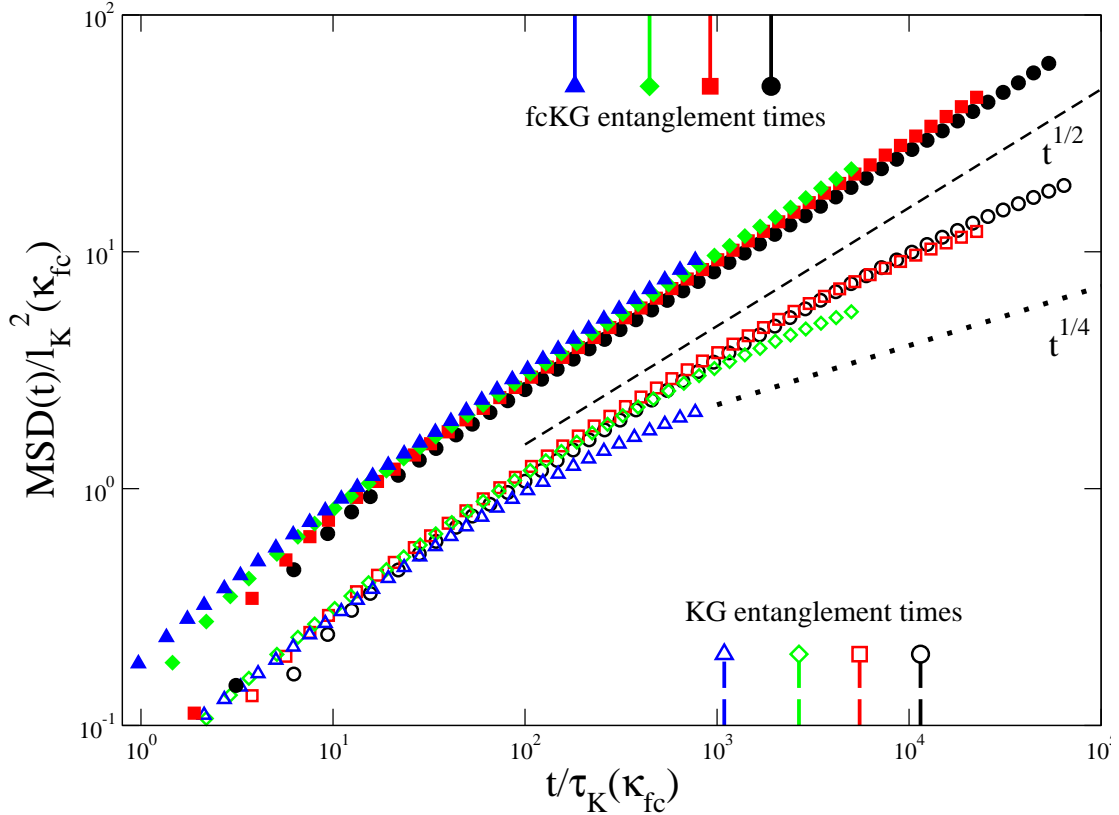}

\caption{\label{fig:msd}Mean-square displacements for the fcKG model (filled
symbols) and KG model (open symbols) for melts with $M\times N_{b}=500\times10.000$
for $\kappa,\kappa_{fc}=-1$ (black circle), $0$ (red box), $1$
(green diamond), and $2$ (blue triangle up). For the KG model the
times were normalized using the time mapping for the corresponding
force-capped KG model to retain their horizontal position. Also indicated
are the entanglement times for the fcKG model and estimated for the
KG model (vertical lines ending a single symbol of the corresponding
model).}
\end{figure}

Figure \ref{fig:msd} shows the mean-square displacements $MSD(t)=\langle[{\bf R}_{i}(0)-{\bf R}_{i}(t)]^{2}\rangle$
of beads for the fcKG and KG models. We observe the expected sub-diffusive
Rouse power law $MSD(t)\sim t^{1/2}$ both above and below the estimated
entanglement for the fcKG model, whereas for the KG model we see the
start of the crossover to a $MSD(t)\sim t^{1/4}$ power law above
the entanglement time. The latter dynamics is due to local reptation
motion inside the tubes. This observation is consistent with our assumption
that the fcKG model produces Rouse dynamics because it allows chains
to pass through each other. Hence the entanglement Rouse time of the
fcKG model is the relevant time for establishing local random walk
structure inside the tube. The mean-square displacements for the KG
models were shifted using the Kuhn times for the corresponding fcKG
models in order to retain the horizontal shift between the curves.
This shift indicates that the dynamics of the full KG model is approximately
a factor of six slower than the fcKG model. We have used this factor
to estimate the entanglement times for the full KG model, which are
also shown in the figure and is in good agreement with the indicated
power law cross-overs.

\subsection{Switching between models}

Using the relations derived above we can fine-grain melt states from
the lattice model to the fcKG model, and later introduce the target
KG model and retaining the desired melt properties through the whole
equilibration process.

After lattice annealing of large scale density fluctuations, the final
lattice polymer melt state is converted into an off-lattice bead-spring
model representation, that can be used as input for the subsequent
Molecular Dynamics simulations. Each lattice chain is translated in
space by a random vector in a cell $[-a/2,a/2]^{3}$, and decorated
with beads to produce the desired target contour length density of
beads. Each bead is furthermore offset by a small random vector in
$[-a/200,a/200]^{3}$ to facilitate energy minimization. This melt
shown in Fig. \ref{fig:visualization} corresponds to a $23\times23\times23$
lattice, and each polymer chain consists of $200$ entanglement blobs.

After the lattice annealing, the Rouse simulation should introduce
random chain structure at progressively larger and larger scales.
Before starting the Rouse simulation, the energy of the final lattice
conformation is minimized with the force-capped  force field. The
melt configuration after $0.1\tau_{e}$ is shown in Fig. \ref{fig:visualization}.
After $0.1\tau_{e}$ the lattice structure is still visible. However,
after $1\tau_{e}$ of Rouse simulation the polymers appears to have
adopted a random walk conformation on the tube scale, and no signs
of the lattice structure remains. After having simulated Rouse dynamics
with the fcKG model for a number of entanglement times $\tau_{e}$,
we expect that the correct chain statistics have been established
on all scales. 

To switch a fcKG melt conformation to the KG force field, it is first
energy minimized using the full WCA pair interaction, but retaining
the fcKG bond potential. Subsequently, the bond potential is changed
to sum of the FENE and WCA potentials and again the energy is minimized.
The resulting melt conformation is then simulated for $5\times10^{4}$
MD steps at $T=1\epsilon$ with the full KG force field to equilibrate
the local bead packing. We denote this procedure the KG warm-up. The
resulting KG configuration is also shown in Fig. \ref{fig:visualization}.
It shows no discernible difference compared to the fcKG melt state
at $1\tau_{e}$.

\section{Characterization of equilibration process\label{sec:Characterization-of-equilibratio}}

Above, we characterized the KG and fcKG models using results for $15$
melt states of $M\times N_{b}=500\times10.000$ i.e. $Z=80,\dots,540$
entanglements for $\kappa=-1,-0.75,-0.50,\cdots,2.25,2.50$. We have
also equilibrated a number of large melts with $M\times N_{b}=1.000\times15.000$
i.e. $Z=200$ but only in the case of  $\kappa=0$. In comparison,
the largest melts produced in Refs. \citep{moreira2015direct,zhang2015communication}
where $1.000$ chains of length $2.000$ beads. We produced eight
melts using the full lattice Hamiltonian described above, five melts
without the incompressibility term, and three melts without the configuration
term. With these variations of the annealing procedure, we can illustrate
why both the incompressibility and configuration terms are required.
The lattice states were simulated with the same Rouse simulation,
but we have also performed the KG warm up at different times during
the Rouse simulation to study how this impacts the resulting KG melts.
Below we will characterize the $1.000\times15.000$ melts states unless
specifying a chain stiffnesses $\kappa$, in which case the observables
are calculated for the $500\times10.000$ melt states.

\begin{figure}
\includegraphics[width=0.45\columnwidth]{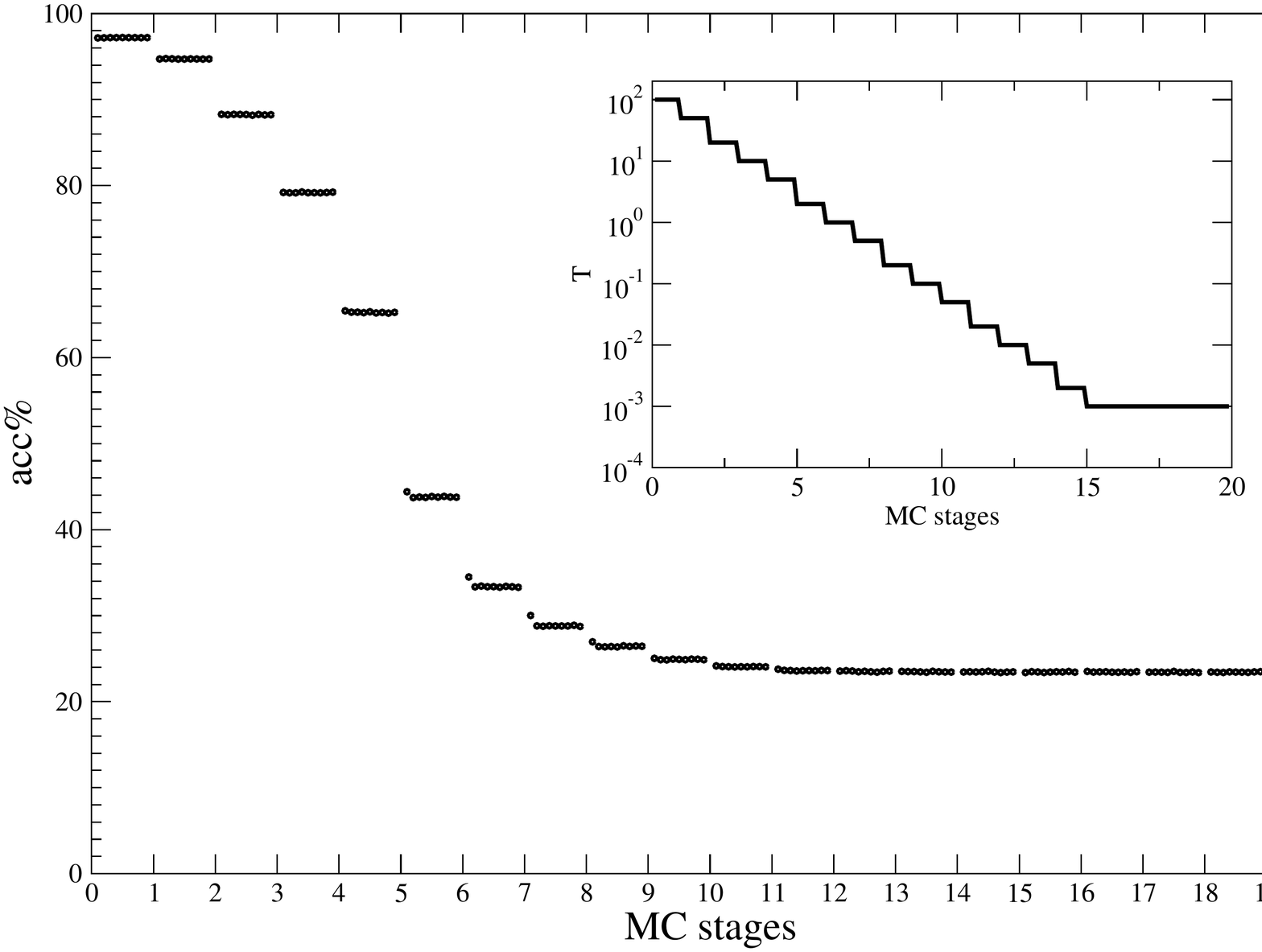}%
\includegraphics[width=0.45\columnwidth]{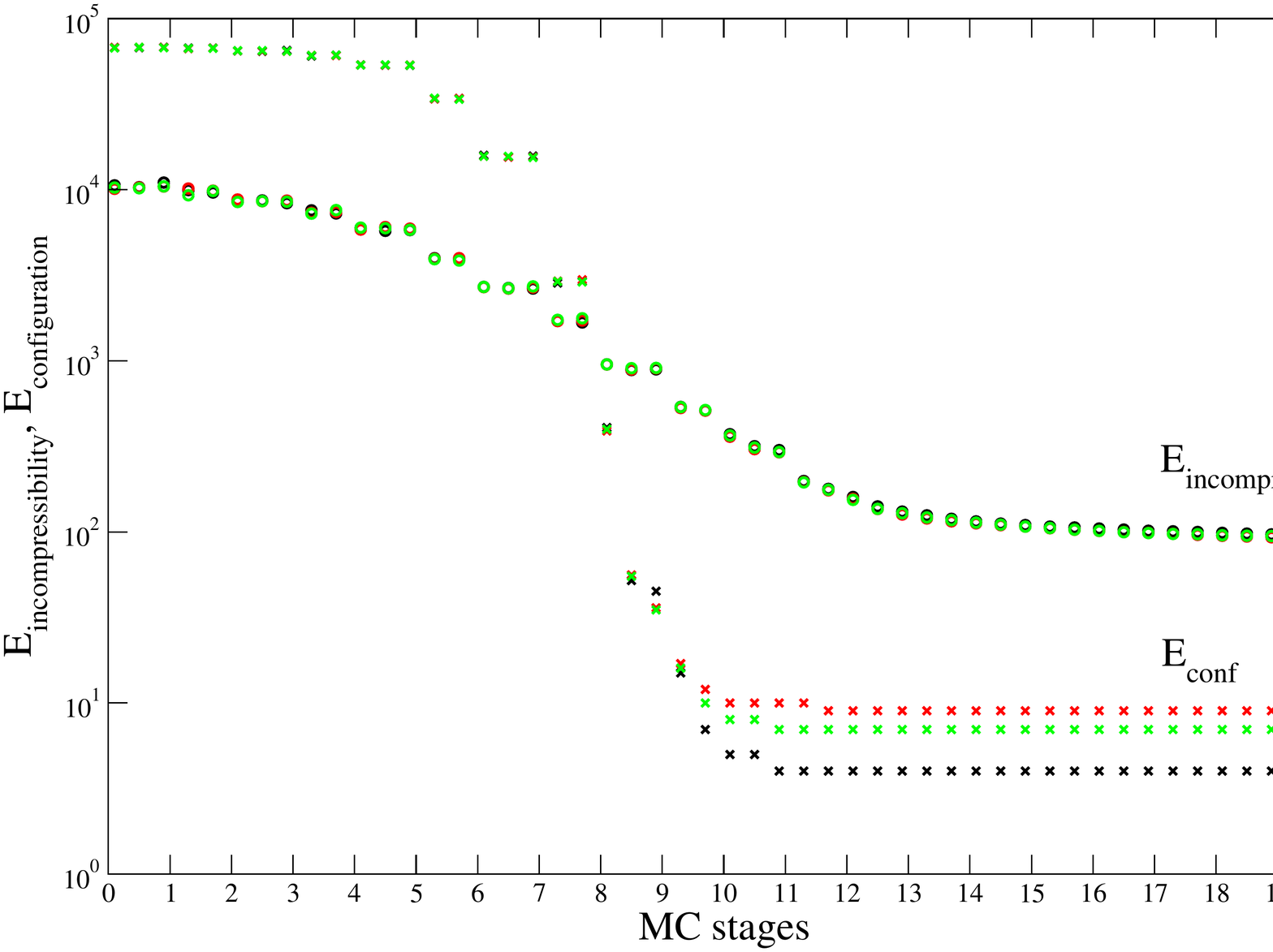}

\caption{Characterization of simulated annealing process showing total acceptance
probability (right) and energy contributions from configuration (circles)
and incompressibility (crosses) terms for three melts (red, green,
black). The insert shows the temperature profile.\label{fig:anneal}}
\end{figure}

Figure \ref{fig:anneal} shows a characterization of the simulated
annealing process. After some experimentation, we chose an annealing
protocol where the temperature is reduced in $20$ annealing stages
from $T=10^{2}$ to $10^{-3}$. At each annealing stage, we attempt
$50$ Monte Carlo (MC) moves per blob in the melt, where we use both
global and local Monte Carlo moves. Above the transition temperature
$T^{*}\sim0.1$, the system rapidly equilibrates and the acceptance
probability shows a clear step structure. Below the transition temperature,
the equilibration slows down considerably and the step like structure
of the acceptance probability is lost. The acceptance rate remains
clearly above $20\%$ even below the transition temperature. This
is primarily due to the end-bridging moves, which are attempted with
$20\%$ probability. The local chain dynamics becomes frozen while
the global chain state remains dynamic, since double bridge moves
are still accepted with even below the transition temperature. Figure
\ref{fig:anneal} also shows the decrease of the energy contributions
from the incompressibility and configuration terms in the lattice
Hamiltonian. The configuration energy contribution drops by about
four orders of magnitude while the incompressibility energy drops
by about two orders of magnitude. Both contributions levels out after
$10-12$ annealing stages. After this time, the three melts have reached
approximately the same energy minimum.

\begin{figure}
\includegraphics[width=0.45\columnwidth]{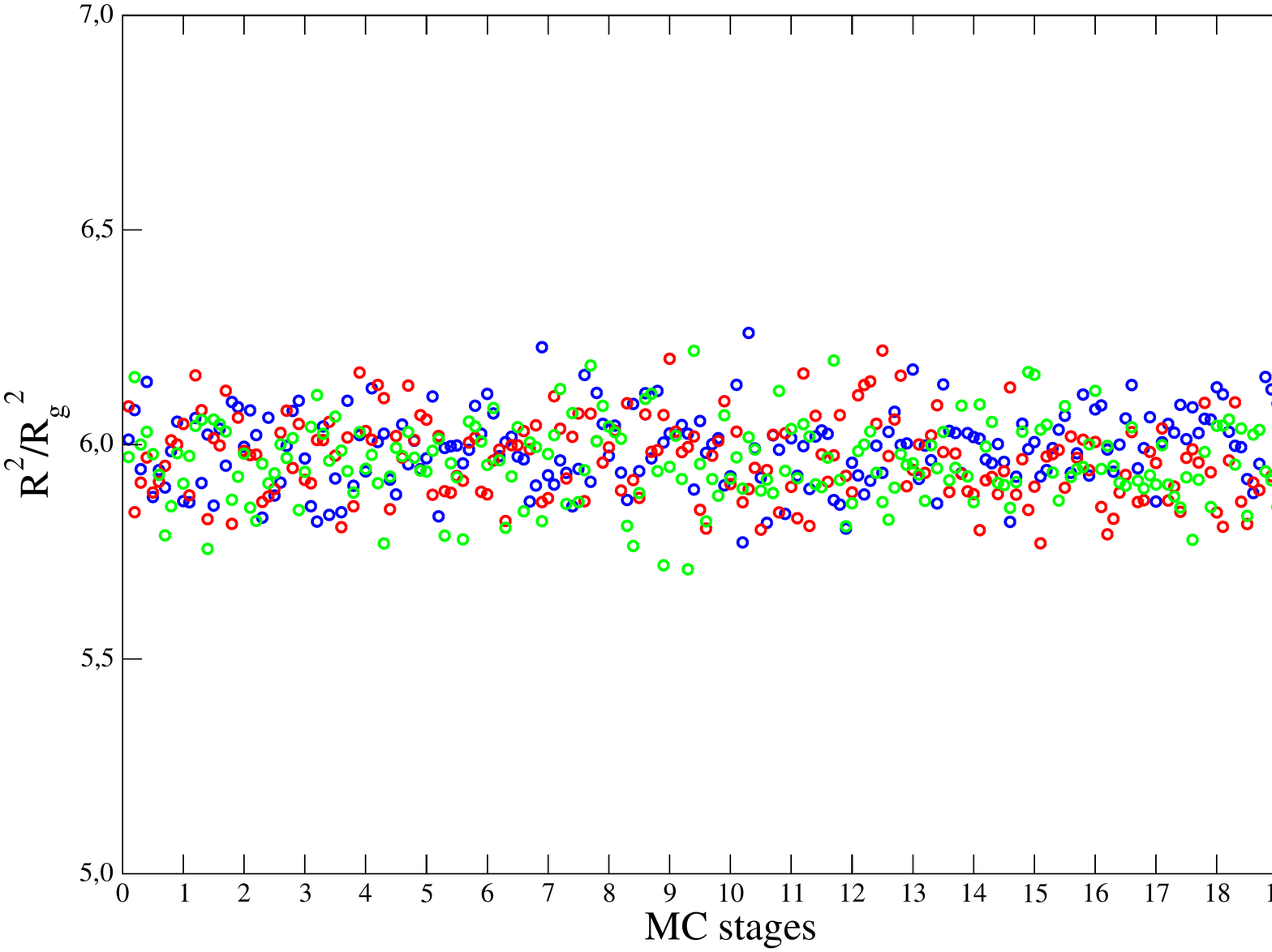}%
\includegraphics[width=0.45\columnwidth]{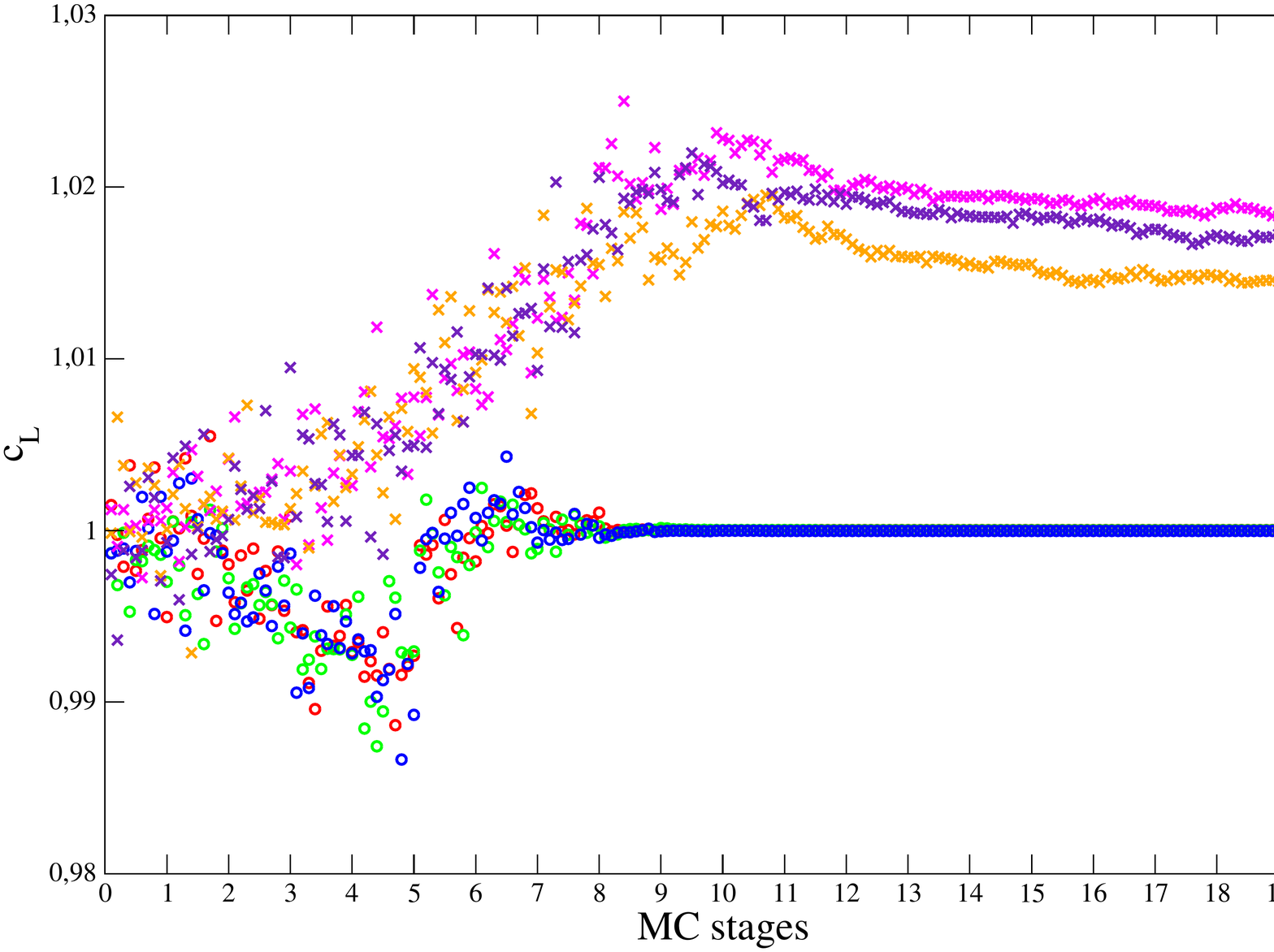}

\caption{\label{fig:characterization}Melt characterization during simulated
annealing showing $\langle R^{2}\rangle/\langle R_{g}^{2}\rangle\approx6$
ratio (left) and chain stiffness $c_{L}$ for lattice melts with the
configuration term (red, green, blue circles) and without the configuration
term in the Hamiltonian (magenta, orange, purple crosses).}
\end{figure}

Figure \ref{fig:characterization} shows how the chain conformations
evolve during the annealing process. We describe the large scale properties
with the ratio of the end-to-end distance and the radius of gyration
which for a random walk should be about six.\citep{flory1989statistical}
We observe that at large scales the melt conformations remain random
walk like during the whole annealing process. Furthermore the scatter
of the curves below the transition temperature again shows that the
MC moves keeps generating new conformations searching for a better
minimum.

The chain stiffness $c_{L}$ characterizes blob chain angle statistics
at the tube scale. This should be unity for random walks where subsequent
steps are statistically uncorrelated. For melts with the configuration
term, this is seen to be the case after some transients around the
transition temperature, however, we see a slight but systematic increase
in the chain stiffness for melts without the configuration term. This
could be either due to the incompressibility constraint acting as
a weak excluded volume even at occupation numbers of $\langle n\rangle\approx19$,
and hence leading to a small degree of swelling. Alternatively, it
is also known that the Flory ideality hypothesis is only approximately
true even for dense melts. The incompressibility constraint leads
to a correlation hole of density fluctuations, which has been shown
to give rise to an effective weakly repulsive intra-molecular interaction.\citep{wittmer2007polymer,wittmer2007intramolecular,beckrich2007intramolecular,meyer2008static,semenov2010bond}
Both these effects lead to swelling, and the severity of the swelling
is likely to depend on the rate at which the simulated annealing process
is quenched. We have opted for adding the additional configuration
term to the lattice Hamiltonian, to ensure that the lattice conformations
show the desired random walk statistics at the smallest scales.

\begin{figure}
\includegraphics[width=0.6\columnwidth]{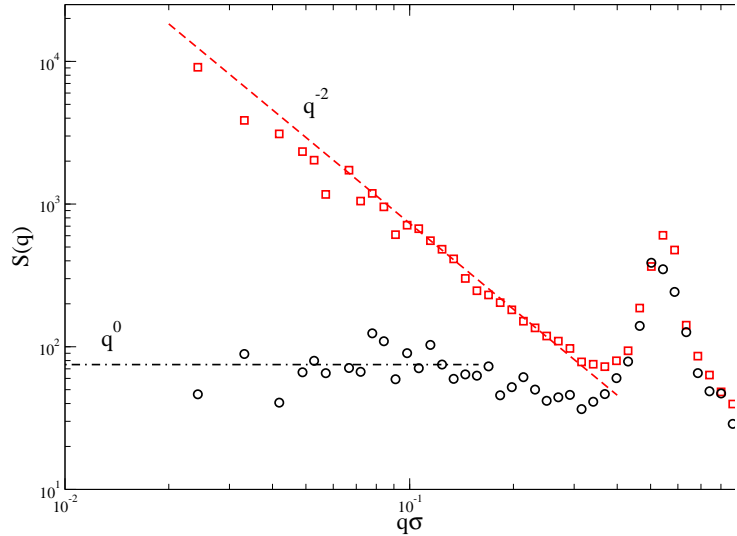}

\caption{\label{fig:Structurefactor-lattice}Structure factor for initial lattice
configurations with density fluctuations (red boxes), and after simulated
annealing with the incompressibility term (black circles). Also shown
are the power laws expected from the density fluctuations and from
incompressibility (hashed and dash-dotted lines)}
\end{figure}

Figure \ref{fig:Structurefactor-lattice} shows the impact of incompressibility
on the lattice melt conformations. We have averaged the structure
factor over several statistically independent melts to improve the
statistics. At the lowest $q$ values, the structure factor characterize
density fluctuations on the scale of the whole simulation domain,
whereas the highest $q$ values reflect density fluctuations on the
scale of individual blobs. The structure factors were calculated for
MD bead-spring melt states and include effects due to random shifts
of chains and beads described above.

For the lattice simulations without the incompressibility term, very
large scale density fluctuations can be seen at large scales, which
follows the predicted power law behavior $S(q)\sim2N_{b}(qR_{g})^{-2}$
(for derivation see the Appendix). This power law reflects the density
fluctuations created by randomly inserting the polymer chains on the
lattice. After annealing, with the the incompressibility term in the
Hamiltonian the large scale density fluctuations are reduced by about
two orders of magnitude, and the resulting structure factor is flat
indicating constant density on all scales as expected for an incompressible
melt. A large peak is seen in both the lattice configurations, this
peak reflects the lattice structure and the position is given by $q_{lattice}=2\pi/a$. 

\begin{figure}
\includegraphics[width=0.6\columnwidth]{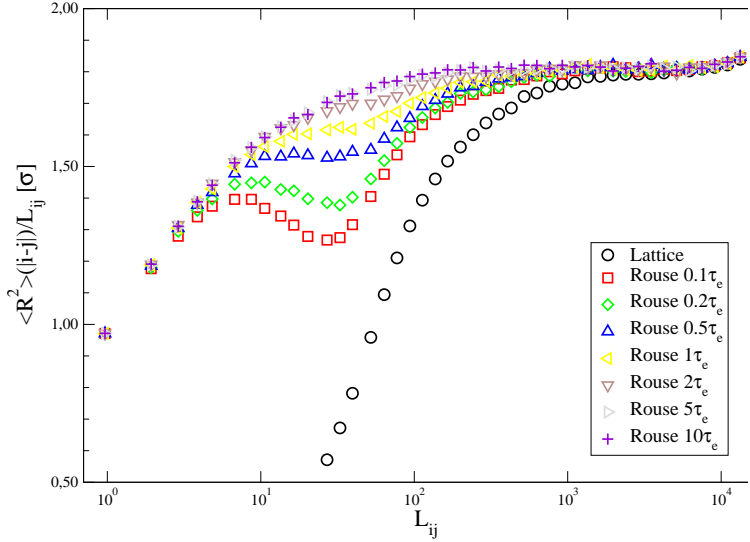}

\caption{\label{fig:Rouse-MSID}Evolution of mean-square internal distances
during equilibration for the initial lattice configuration (black
circle), and after $0.1\tau_{e}$, $0.2\tau_{e},0.5\tau_{e},1\tau_{e}$,
$2\tau_{e}$, $5\tau_{e}$, $10\tau_{e}$, $20\tau_{e}$ of Rouse
simulation time with the fcKG for $\kappa=0$. }
\end{figure}

Figure \ref{fig:Rouse-MSID} shows the evolution of single chain conformations
characterized by their mean-square internal distances (MSID), which
are defined by $MSID(L_{ij})=\langle({\bf R}{}_{i}-{\bf R}{}_{j})^{2}\rangle/L_{ij}$
where ${\bf R}{}_{i}$ is the position of the $i$'th bead on a chain,
and $L_{ij}=l_{b}|i-j|$ denotes the chemical contour length between
the two beads. For large chemical distances the MSID converges to
the Kuhn length, whereas for neighboring monomers it is identical
to the bond length $l_{b}$. Between these limits it characterize
the local effects of the chain stiffness.

The evolution of the chain statistics during the Rouse simulation
is shown in Fig. \ref{fig:Rouse-MSID}. The final state from the lattice
simulation matches the large scale chain statistics by construction,
but shows strong compression at all length scales below the tube diameter,
which is an expected lattice artifact. After energy minimization and
a brief simulation, the bond distance agrees with the KG model, but
chains are stretched at very short scales, and compressed at scales
all the way to the tube scale. During the Rouse simulation, the chain
statistics is progressively equilibrated at intermediate scales such
that the desired chain statistics is established on all length scales.
In the initial lattice configuration all the beads are lying on a
straight line and hence we approach the equilibrium chain statistics
from below, whereas in the approach of Auhl et al.\citep{auhl2003equilibration},
their push off produced a peak in the MSID that is due to local chain
stretching due to density fluctuations, which was mitigated by the
introduction of a pre-packing procedure. Here our fcKG model has been
designed to perform this pre-packing on scales below the tube diameter
during the Rouse simulation.

\begin{figure}
\includegraphics[width=0.7\columnwidth]{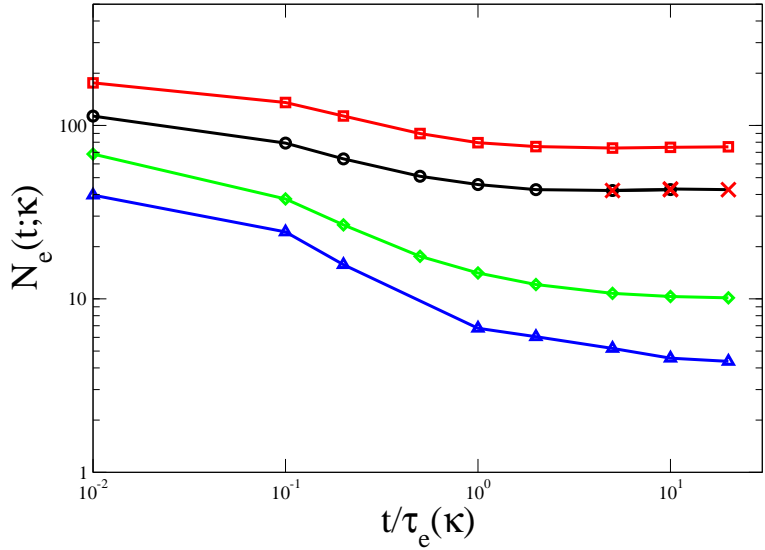}

\caption{\label{fig:EntanglementLength}Topological evolution of the melt during
the Rouse simulation. The entanglement lengths of the force-capped
KG model melts for $\kappa=-1\epsilon$ (red box), $0\epsilon$ (black
circle), $1.5\epsilon$ (green diamond), and $2.5\epsilon$ (blue
triangle up). Also shown are the entanglement length of melts after
the KG warm up for three different times along the Rouse simulation
for $\kappa=0\epsilon$ (red crosses).}
\end{figure}

The MSID is a single chain observable, we can also take melt configurations
at various times along the Rouse dynamics simulation and submit them
to PPA analysis to estimate the topological evolution of the melt.
The entanglement length has been shown to be quite sensitive to the
equilibration procedure, since chain stretching during equilibration
of badly prepared samples artificially increase the entanglement density.\citep{hoy2005eep}
The result is shown in Fig. \ref{fig:EntanglementLength}. The entanglement
length is seen to systematically decrease towards the equilibrium
entanglement length after about one entanglement time of Rouse dynamics
independently of chain stiffness. During the Rouse simulation chains
can pass through each other, however, during the PPA the topological
structure is frozen. Hence the figure shows the growth of the entanglement
density due to the random chain structure that is gradually introduced
by the Rouse dynamics of the fcKG model. The figure also shows that
the initial lattice states produce a completely wrong entanglement
density, hence any attempt to equilibrate it with a topology preserving
chain model would fail. Finally, the entanglement length of the three
conformations after the KG warm up are in excellent agreement with
that of the Rouse simulations, as expected the KG warm up does not
change the topological structure of the melts.

\begin{figure}
\includegraphics[width=0.6\columnwidth]{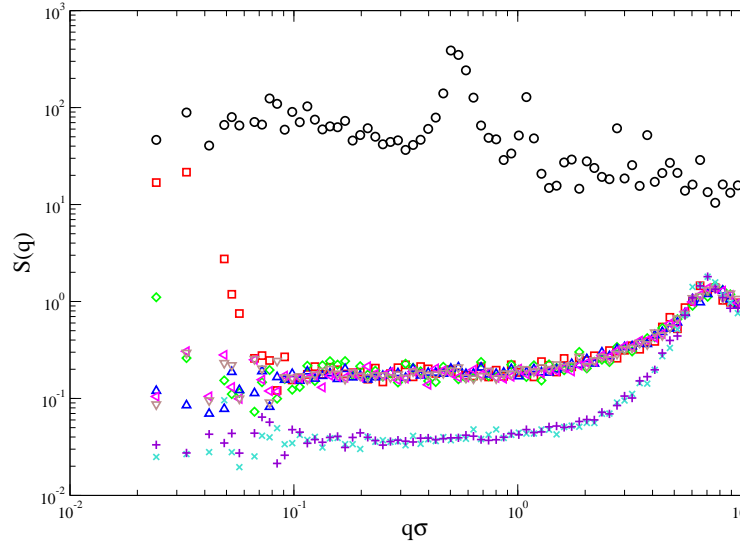}

\caption{\label{fig:Structurefactor}Evolution of structure factor during equilibration
process. Structure factor for the annealed lattice (black circle),
force-capped KG model after $0.1\tau_{e}$ (red box), $0.5\tau_{e}$
(green diamond), $1\tau_{e}$ (blue triangle up), $5\text{\ensuremath{\tau}}_{e}$
(magenta triangle right), $10\tau_{e}$ (brown triangle down) of simulation,
and finally after KG warm for configurations after $5\tau_{e}$ (turquoise
cross) or $10\tau_{e}$ (violet plus) of Rouse simulation.}
\end{figure}

The structure factor during the Rouse simulation and after the KG
warm up is shown in Fig. \ref{fig:Structurefactor}. The structure
factor measures density fluctuations and when constant allows us to
estimate the compressibility of the melt (for derivation see the Appendix).
We see that the melt compressibility rapidly decrease by about three
orders of magnitude when lattice melt states are equilibrated with
the fcKG model. Residual large scale density fluctuations are still
observable at $0.1\tau_{e}$, but after $1\tau_{e}$ density fluctuations
are absent on all scales. After the KG warm up, the compressibility
is further reduced by about one order of magnitude on all scales.
This is also seen to be independent of when we perform the KG warm
up. A rising tendency of the structure factors are observed at large
$q$ values, which is due to the first Bragg peak at $2\pi/\sigma$
and is due to local liquid like bead packing. The structure factors
for melts with varying stiffness show similar behavior (data not shown).

\begin{figure}
\includegraphics[width=0.6\columnwidth]{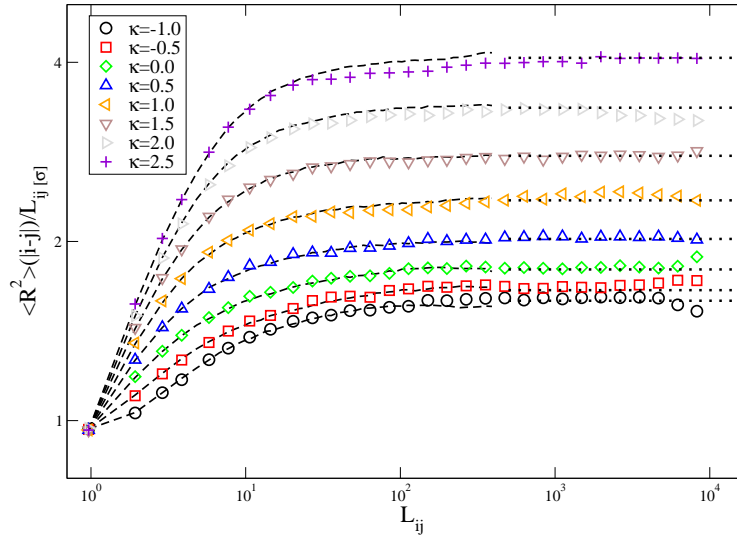}

\caption{\label{fig:MDIkappa}Mean-square internal distances of equilibrated
melts (coloured symbols) compared to brute force equilibrated melts
(dashed black lines) and Kuhn length (dotted black line) for varying
stiffness.}
\end{figure}

Figure \ref{fig:MDIkappa} compares the MSID for different stiffness
of rapidly equilibrated long melts and brute force equilibrated shorter
melts used to estimate the Kuhn lengths. The Rouse simulations were
performed for $10\tau_{e}$ when $\kappa<1.5\epsilon$, $20\tau_{e}$
for $\kappa=1.5\epsilon$, $30\tau_{e}$ for $\kappa=2.0\epsilon$,
and $40\tau_{e}$ for $\kappa=2.5\epsilon$. Note that since the entanglement
time drops rapidly with increasing chain stiffness, the actual run
time is largest for the equilibration of the most flexible melts with
$\kappa=-1$, which require about a factor of three longer simulation
time than the stiffest melts. The chain statistics shown in the figure
is in good agreement with the brute force equilibrated melts at short
scales, furthermore all the melts levels off to the expected plateau
given by the Kuhn length at large scales. A small dip is seen for
the two stiffest melts shown in the figure for an intermediate length
scale $L_{ij}\approx100$. Perhaps the stiffest melts are locally
nematically ordered, in which case the energy barrier for chain interpenetration
could be larger than expected and hence explain why we apparently
need to run the simulation for longer than expected from the entanglement
time expected from Rouse dynamics. Clearly, the MSID is the measure
that is the slowest to converge to the equilibrium since the bulk
properties and collective mesoscopic properties measured by the structure
factor and melt entanglement length have already reached their equilibrium
values after $1\tau_{e}$.

\section{Conclusions\label{sec:Conclusions}}

We have shown how to equilibrate huge model polymer melts in three
simple stages for Kremer-Grest polymer melts\citep{grest1986molecular}
of varying chain stiffness. Firstly, density fluctuations are annealed
on scales above the tube scale using simulated annealing with a lattice
polymer model. Secondly, with a Molecular Dynamics simulation of a
force capped Kremer-Grest (fcKG) polymer model, we simulate the Rouse
dynamics\citep{rouse1953theory} and introduce the desired chain structure
on the tube scale and below while preventing the growth of density
fluctuations. Finally, we perform a fast warm up to the Kremer-Grest
force field to establish the correct local bead packing. We have characterized
the involved models in order to transfer melt states between them
for varying chain stiffnesses. By measuring the Rouse friction of
the fcKG model, we have also estimated the simulation time required
for the equilibration of chain structure inside the tube, which was
shown to be strongly dependent on chain stiffness.

We have also characterized and validated the equilibration process
in terms of 1) single chain observables such as mean-square internal
distances, 2) collective mesoscopic melt properties such as the evolution
of the entanglement length during the Rouse dynamics, and 3) bulk
melt density fluctuations in terms of structure factors. We have demonstrated
the convergence of these observables to their equilibrium values for
varying chain stiffness.

The main requirement of a equilibration process is computational performance.
Here we have equilibrated $15$ melts of $500$ chains with $10.000$
beads each for varying stiffness, and several melts of $1.000$ chains
of $15.000$ beads each for varying lattice annealing parameters.
For the latter melts, the lattice annealing of density fluctuations
takes about $3$ days computer time using a single core on a standard
laptop ($72$ core hours). Rouse simulation with the fcKG model for
$10\tau_{e}$ takes about 2 days on 4 ABACUS2 nodes\citep{AbacusHardware},
i.e. $4.600$ core hours of compute time. Finally introducing the
full KG model, requires about 2 hours on 4 nodes, i.e. $200$ core
hours. Moreira\citep{moreira2015direct} equilibrated $1.000\times2.000$
melts using $3.500$ core hours for pre-packing (equivalent to our
lattice annealing), and $3.800$ core hours for subsequent warm up.
Zhang et al. \citep{zhang2014equilibration} equilibrated similar
sized melts but using a multiscale method that required $1.600$ core
hours. Scaling these numbers to a standard melt of one million beads,
the method of Moreira et al. would require $3.600$ core hours, the
method of Zhang et al. would requires $800$ core hours, while our
method would require $600$ core hours. This could be further optimized
e.g. the choice of the force cap is entirely serendipitous, and we
use the standard values of the Kremer-Grest polymer model, which could
be optimized further.

The present method is essentially independent of e.g. chain length
and the polymer structure. These only impacts the lattice annealing
stage of the equilibration procedure, which is the fastest part of
the equilibration process. The Rouse simulation and KG warm up are
completely independent of the chain structure and composition. Hence
the present method can directly be used to equilibrate e.g. polymer
melts of stars or mixtures of different polymer structures. Furthermore,
the equilibrated KG melt configurations produced by the present approach
can be fine-grained further to act as starting points for atomistic
simulations of polymer melts. 

With simple and computationally efficient equilibration approaches
such as the one presented here, access to well equilibrated melts
for studies of material properties is no longer a computational limitation,
rather the limitation is the computational effort required to estimate
the material properties of these huge systems.

\section{Acknowledgements}

Computation/simulation for the work described in this paper was supported
by the DeiC National HPC Center at University of Southern Denmark.

\section{Appendix}

Below we derive predictions for the structure factor due to the density
fluctuations created by randomly inserting polymers in the simulation
box, and the structure factor after equilibration of density fluctuations
and its relation to the compressibility.

Defining the microscopic density as microscopic density field $\rho({\bf R})=\sum_{j=1}^{M}\sum_{k=1}^{N}\delta({\bf R}-{\bf R}_{jk})$,
where $\delta$ denotes the Dirac-delta function. We can then recognize
the Fourier transform of the density field is $\rho({\bf q})=\sum_{j=1}^{M}\sum_{k=1}^{N_{b}}\exp(i{\bf q}\cdot{\bf R}_{jk})$,
such that the structure factor can be written

\begin{equation}
S(q)=\left(N_{b}M\right)^{-1}\langle\rho(-{\bf q})\rho({\bf q})\rangle\label{eq:structurefactor_density}
\end{equation}

To derive the structure factor after equilibration of density fluctuation,
we start by expressing the structure factor in terms of spatially
varying densities. From the right hand side of eq. \ref{eq:structurefactor_density}
we get

\[
S({\bf q})=\left\langle \int\mbox{d}{\bf R}_{1}{\bf \mbox{d}}{\bf R}_{2}\rho({\bf R}_{1})\rho({\bf R}_{2})\exp(i{\bf q}\cdot({\bf R}_{1}-{\bf R}_{2}))\right\rangle 
\]

\[
=\int\mbox{d}{\bf R}_{1}{\bf \mbox{d}}{\bf R}_{2}\exp(i{\bf q}\cdot({\bf R}_{1}-{\bf R}_{2}))\left\langle \delta\rho({\bf R}_{1})\delta\rho({\bf R}_{2})\right\rangle 
\]

\begin{equation}
=\int\mbox{d}{\bf R}\exp(i{\bf q}\cdot{\bf R})\left\langle \delta\rho(0)\delta\rho({\bf R})\right\rangle \label{eq:fouriertransformdensfluc}
\end{equation}
where in the second equation we have replaced $\rho({\bf R})\rightarrow\delta\rho({\bf R})=\rho({\bf R})-\langle\rho\rangle$.
The constant average density gives rise to a contribution proportional
to a Dirac-delta function, which can be neglected for $q>0$. In the
third equation, we have furthermore assumed translational invariance. 

Lets assume a local Hamiltonian for density fluctuations $H(\delta\rho)=\frac{1}{2\chi\langle\rho\rangle}\delta\rho^{2}$
for a particular position analogous to a site in the lattice model.
The Boltzmann probability of a given density fluctuation is given
by $P(\delta\rho)\propto\exp(-H/kT)$, and hence $\left\langle \delta\rho^{2}\right\rangle =\chi\langle\rho\rangle kT$
by the equipartition theorem. Assuming that the density fluctuations
at different sites are statistically independent, which in practice
is valid for sufficiently large distances, i.e. small values of $q$.
The density fluctuation correlation function is then given by $\left\langle \delta\rho(0)\delta\rho({\bf R})\right\rangle =\chi\langle\rho\rangle kT\delta({\bf R})$.
Inserting this in eq. \ref{eq:fouriertransformdensfluc}, we obtain
the prediction that the structure factor is independent of $q$ and
proportional to the compressibility

\[
S(q)=\chi\langle\rho\rangle kT.
\]

We can also predict the structure factor for polymers randomly inserted
into the simulation domain using the approach in Refs. \citep{svaneborg2012formalism,svaneborg2012formalism2}.
By introducing an origin of the coordinate system for each polymer
${\bf R}_{j}^{o}$ e.g. one of its ends, we can rewrite eq. \ref{eq:structurefactorpoint}
as

\[
S(q)=\left(N_{b}M\right)^{-1}\sum_{k_{1}=1}^{N_{b}}\sum_{k_{2}=1}^{N_{b}}\left[\sum_{j=1}^{M}\left\langle \exp(i{\bf q}\cdot\left({\bf R}_{jk_{1}}-{\bf R}_{jk_{2}}\right))\right\rangle \right.
\]

\[
\left.+\sum_{\begin{array}{c}
j_{1},j_{2}=1\\
j_{1}\neq j_{2}
\end{array}}^{M}\left\langle \exp(i{\bf q}\cdot\left[\left({\bf R}_{j_{1}k_{1}}-{\bf R}_{j_{1}}^{o}\right)-\left({\bf R}_{j_{2}k_{2}}-{\bf R}_{j_{2}}^{o}\right)+\left({\bf R}_{j_{1}}^{o}-{\bf R}_{j_{2}}^{o}\right)\right])\right\rangle \right],
\]
here the first term describes the single polymer scattering due to
pairs of scattering sites on the same polymer, while the second term
is the interference contribution between scattering sites on different
polymers. Configurations of different polymers are generated independently
of each other, and the starting points of the polymers are chosen
randomly. Hence the three terms in parenthesis in the second exponential
are sampled from statistically independent distributions. Having noted
this, the average of the interference contribution factorize exactly
into a product of three averages, where the first and third only depend
on single chain statistics, while the second term only depends on
the distance distribution between randomly chosen points:

\[
\left\langle \exp(i{\bf q}\cdot\left({\bf R}_{j_{1}k_{1}}-{\bf R}_{j_{1}}^{o}\right))\right\rangle _{j_{1},k}\left\langle \exp(i{\bf q}\cdot\left({\bf R}_{j_{1}}^{o}-{\bf R}_{j_{2}}^{o}\right))\right\rangle _{j_{1},j_{2}}\left\langle \exp(-i{\bf q}\cdot\left({\bf R}_{j_{2}k_{2}}-{\bf R}_{j_{2}}^{o}\right))\right\rangle _{j_{2},k_{2}}.
\]

For notational simplicity, we can identify the average form factor
as $F({\bf q})=\left\langle \exp(i{\bf q}\cdot\left({\bf R}_{jk_{1}}-{\bf R}_{jk_{2}}\right))\right\rangle _{j,k_{1},k_{2}}$
with the average single polymer scattering, $A({\bf q})=\left\langle \exp(i{\bf q}\cdot\left({\bf R}_{jk_{1}}-{\bf R}_{j}^{o}\right))\right\rangle _{j,k_{1}}$
with the average form factor amplitude relative to an end, and the
average phase factor between different ends $\Psi({\bf q})=\left\langle \exp(i{\bf q}\cdot\left[\left({\bf R}_{j_{1}}^{o}-{\bf R}_{j_{2}}^{o}\right)\right])\right\rangle _{j_{1}\neq j_{2}}$.
Note all these factors are normalized as $F({\bf q})=A({\bf q})=\Psi({\bf q})\rightarrow1$
for ${\bf q}\rightarrow0$. With these simplifications, the structure
factor reduce to the much shorter expression

\[
S({\bf q})=N_{b}F({\bf q})+(M-1)N_{b}A({\bf q})\Psi({\bf q})A(-{\bf q}).
\]
this expression is exact and was derived without any assumptions as
to the detailed structure of the objects inserted in the simulation
domain and only results from the assumption of statistical independence
of intrnal conformations and positions of the objects.\citep{svaneborg2012formalism}
For polymers modeled as ideal random walks, the expressions for the
form factor and form factor amplitude are well known\citep{debye1947formfactor,hammouda1992formfactoramplitude}:

\[
F(q)=\frac{2\left[\exp(-q^{2}R_{g}^{2})-1+q^{2}R_{g}^{2}\right]}{q^{4}R_{g}^{4}}\sim\frac{2}{q^{2}R_{g}^{2}},
\]

\[
A(q)=\frac{1-\exp(-q^{2}R_{g}^{2})}{q^{2}R_{g}^{2}}\sim\frac{1}{q^{2}R_{g}^{2}},
\]
since random walks on average are isotropic, these functions only
depends on the magnitude of the scattering vector $q=|{\bf q}|$.
The asymptotic behavior is realized for $qR_{g}\gg1$, where $R_{g}=l_{K}^{2}N_{k}/6$
is the radius of gyration of the polymer. Furthermore, because polymers
pairs are placed randomly in the box their starting positions are
statistically independent, hence $\Psi({\bf q})=1$. Hence for randomly
inserted polymers, the asymptotic behavior of the structure factor
describing the resulting density fluctuation correlations is

\[
S(q)\sim\frac{2N_{b}}{q^{2}R_{g}^{2}}.
\]

\bibliographystyle{apsrev4-1}

\begin{thebibliography}{69}%
\makeatletter
\providecommand \@ifxundefined [1]{%
 \@ifx{#1\undefined}
}%
\providecommand \@ifnum [1]{%
 \ifnum #1\expandafter \@firstoftwo
 \else \expandafter \@secondoftwo
 \fi
}%
\providecommand \@ifx [1]{%
 \ifx #1\expandafter \@firstoftwo
 \else \expandafter \@secondoftwo
 \fi
}%
\providecommand \natexlab [1]{#1}%
\providecommand \enquote  [1]{``#1''}%
\providecommand \bibnamefont  [1]{#1}%
\providecommand \bibfnamefont [1]{#1}%
\providecommand \citenamefont [1]{#1}%
\providecommand \href@noop [0]{\@secondoftwo}%
\providecommand \href [0]{\begingroup \@sanitize@url \@href}%
\providecommand \@href[1]{\@@startlink{#1}\@@href}%
\providecommand \@@href[1]{\endgroup#1\@@endlink}%
\providecommand \@sanitize@url [0]{\catcode `\\12\catcode `\$12\catcode
  `\&12\catcode `\#12\catcode `\^12\catcode `\_12\catcode `\%12\relax}%
\providecommand \@@startlink[1]{}%
\providecommand \@@endlink[0]{}%
\providecommand \url  [0]{\begingroup\@sanitize@url \@url }%
\providecommand \@url [1]{\endgroup\@href {#1}{\urlprefix }}%
\providecommand \urlprefix  [0]{URL }%
\providecommand \Eprint [0]{\href }%
\providecommand \doibase [0]{http://dx.doi.org/}%
\providecommand \selectlanguage [0]{\@gobble}%
\providecommand \bibinfo  [0]{\@secondoftwo}%
\providecommand \bibfield  [0]{\@secondoftwo}%
\providecommand \translation [1]{[#1]}%
\providecommand \BibitemOpen [0]{}%
\providecommand \bibitemStop [0]{}%
\providecommand \bibitemNoStop [0]{.\EOS\space}%
\providecommand \EOS [0]{\spacefactor3000\relax}%
\providecommand \BibitemShut  [1]{\csname bibitem#1\endcsname}%
\let\auto@bib@innerbib\@empty
\bibitem [{\citenamefont {Snijkers}\ \emph {et~al.}(2015)\citenamefont
  {Snijkers}, \citenamefont {Pasquino}, \citenamefont {Olmsted},\ and\
  \citenamefont {Vlassopoulos}}]{snijkers2015perspectives}%
  \BibitemOpen
  \bibfield  {author} {\bibinfo {author} {\bibfnamefont {F.}~\bibnamefont
  {Snijkers}}, \bibinfo {author} {\bibfnamefont {R.}~\bibnamefont {Pasquino}},
  \bibinfo {author} {\bibfnamefont {P.}~\bibnamefont {Olmsted}}, \ and\
  \bibinfo {author} {\bibfnamefont {D.}~\bibnamefont {Vlassopoulos}},\
  }\href@noop {} {\bibfield  {journal} {\bibinfo  {journal} {J. Phys. Condens.
  Matter.}\ }\textbf {\bibinfo {volume} {27}},\ \bibinfo {pages} {473002}
  (\bibinfo {year} {2015})}\BibitemShut {NoStop}%
\bibitem [{\citenamefont {Padding}\ and\ \citenamefont
  {Briels}(2011)}]{padding2011systematic}%
  \BibitemOpen
  \bibfield  {author} {\bibinfo {author} {\bibfnamefont {J.}~\bibnamefont
  {Padding}}\ and\ \bibinfo {author} {\bibfnamefont {W.}~\bibnamefont
  {Briels}},\ }\href@noop {} {\bibfield  {journal} {\bibinfo  {journal} {J.
  Phys. Condens. Matter}\ }\textbf {\bibinfo {volume} {23}},\ \bibinfo {pages}
  {233101} (\bibinfo {year} {2011})}\BibitemShut {NoStop}%
\bibitem [{\citenamefont {Li}\ \emph {et~al.}(2013)\citenamefont {Li},
  \citenamefont {Abberton}, \citenamefont {Kr{\"o}ger},\ and\ \citenamefont
  {Liu}}]{li2013challenges}%
  \BibitemOpen
  \bibfield  {author} {\bibinfo {author} {\bibfnamefont {Y.}~\bibnamefont
  {Li}}, \bibinfo {author} {\bibfnamefont {B.~C.}\ \bibnamefont {Abberton}},
  \bibinfo {author} {\bibfnamefont {M.}~\bibnamefont {Kr{\"o}ger}}, \ and\
  \bibinfo {author} {\bibfnamefont {W.~K.}\ \bibnamefont {Liu}},\ }\href@noop
  {} {\bibfield  {journal} {\bibinfo  {journal} {Polymers}\ }\textbf {\bibinfo
  {volume} {5}},\ \bibinfo {pages} {751} (\bibinfo {year} {2013})}\BibitemShut
  {NoStop}%
\bibitem [{\citenamefont {Everaers}\ \emph {et~al.}(2004)\citenamefont
  {Everaers}, \citenamefont {Sukumaran}, \citenamefont {Grest}, \citenamefont
  {Svaneborg}, \citenamefont {Sivasubramanian},\ and\ \citenamefont
  {Kremer}}]{everaers2004rheology}%
  \BibitemOpen
  \bibfield  {author} {\bibinfo {author} {\bibfnamefont {R.}~\bibnamefont
  {Everaers}}, \bibinfo {author} {\bibfnamefont {S.~K.}\ \bibnamefont
  {Sukumaran}}, \bibinfo {author} {\bibfnamefont {G.~S.}\ \bibnamefont
  {Grest}}, \bibinfo {author} {\bibfnamefont {C.}~\bibnamefont {Svaneborg}},
  \bibinfo {author} {\bibfnamefont {A.}~\bibnamefont {Sivasubramanian}}, \ and\
  \bibinfo {author} {\bibfnamefont {K.}~\bibnamefont {Kremer}},\ }\href@noop {}
  {\bibfield  {journal} {\bibinfo  {journal} {Science}\ }\textbf {\bibinfo
  {volume} {303}},\ \bibinfo {pages} {823} (\bibinfo {year}
  {2004})}\BibitemShut {NoStop}%
\bibitem [{\citenamefont {Doi}\ and\ \citenamefont
  {Edwards}(1986)}]{DoiEdwards_86}%
  \BibitemOpen
  \bibfield  {author} {\bibinfo {author} {\bibfnamefont {M.}~\bibnamefont
  {Doi}}\ and\ \bibinfo {author} {\bibfnamefont {S.~F.}\ \bibnamefont
  {Edwards}},\ }\href@noop {} {\emph {\bibinfo {title} {The Theory of Polymer
  Dynamics}}}\ (\bibinfo  {publisher} {Claredon Press},\ \bibinfo {address}
  {Oxford},\ \bibinfo {year} {1986})\BibitemShut {NoStop}%
\bibitem [{\citenamefont {de~Gennes}(1971)}]{de1971reptation}%
  \BibitemOpen
  \bibfield  {author} {\bibinfo {author} {\bibfnamefont {P.-G.}\ \bibnamefont
  {de~Gennes}},\ }\href@noop {} {\bibfield  {journal} {\bibinfo  {journal} {J.
  Chem. Phys.}\ }\textbf {\bibinfo {volume} {55}},\ \bibinfo {pages} {572}
  (\bibinfo {year} {1971})}\BibitemShut {NoStop}%
\bibitem [{\citenamefont {de~Gennes}(1976)}]{de1976dynamics}%
  \BibitemOpen
  \bibfield  {author} {\bibinfo {author} {\bibfnamefont {P.-G.}\ \bibnamefont
  {de~Gennes}},\ }\href@noop {} {\bibfield  {journal} {\bibinfo  {journal}
  {Macromolecules}\ }\textbf {\bibinfo {volume} {9}},\ \bibinfo {pages} {587}
  (\bibinfo {year} {1976})}\BibitemShut {NoStop}%
\bibitem [{\citenamefont {Klein}(1978)}]{klein1978evidence}%
  \BibitemOpen
  \bibfield  {author} {\bibinfo {author} {\bibfnamefont {J.}~\bibnamefont
  {Klein}},\ }\href@noop {} {\bibfield  {journal} {\bibinfo  {journal}
  {Nature}\ }\textbf {\bibinfo {volume} {271}},\ \bibinfo {pages} {143}
  (\bibinfo {year} {1978})}\BibitemShut {NoStop}%
\bibitem [{\citenamefont {Zamponi}\ \emph {et~al.}(2005)\citenamefont
  {Zamponi}, \citenamefont {Monkenbusch}, \citenamefont {Willner},
  \citenamefont {Wischnewski}, \citenamefont {Farago},\ and\ \citenamefont
  {Richter}}]{zamponi2005contour}%
  \BibitemOpen
  \bibfield  {author} {\bibinfo {author} {\bibfnamefont {M.}~\bibnamefont
  {Zamponi}}, \bibinfo {author} {\bibfnamefont {M.}~\bibnamefont
  {Monkenbusch}}, \bibinfo {author} {\bibfnamefont {L.}~\bibnamefont
  {Willner}}, \bibinfo {author} {\bibfnamefont {A.}~\bibnamefont
  {Wischnewski}}, \bibinfo {author} {\bibfnamefont {B.}~\bibnamefont {Farago}},
  \ and\ \bibinfo {author} {\bibfnamefont {D.}~\bibnamefont {Richter}},\
  }\href@noop {} {\bibfield  {journal} {\bibinfo  {journal} {Europhys. Lett.}\
  }\textbf {\bibinfo {volume} {72}},\ \bibinfo {pages} {1039} (\bibinfo {year}
  {2005})}\BibitemShut {NoStop}%
\bibitem [{\citenamefont {Pearson}\ and\ \citenamefont
  {Helfand}(1984)}]{pearson1984viscoelastic}%
  \BibitemOpen
  \bibfield  {author} {\bibinfo {author} {\bibfnamefont {D.~S.}\ \bibnamefont
  {Pearson}}\ and\ \bibinfo {author} {\bibfnamefont {E.}~\bibnamefont
  {Helfand}},\ }\href@noop {} {\bibfield  {journal} {\bibinfo  {journal}
  {Macromolecules}\ }\textbf {\bibinfo {volume} {17}},\ \bibinfo {pages} {888}
  (\bibinfo {year} {1984})}\BibitemShut {NoStop}%
\bibitem [{\citenamefont {Brown}\ \emph {et~al.}(1994)\citenamefont {Brown},
  \citenamefont {Clarke}, \citenamefont {Okuda},\ and\ \citenamefont
  {Yamazaki}}]{BrownJCP1990}%
  \BibitemOpen
  \bibfield  {author} {\bibinfo {author} {\bibfnamefont {D.}~\bibnamefont
  {Brown}}, \bibinfo {author} {\bibfnamefont {J.~H.~R.}\ \bibnamefont
  {Clarke}}, \bibinfo {author} {\bibfnamefont {M.}~\bibnamefont {Okuda}}, \
  and\ \bibinfo {author} {\bibfnamefont {T.}~\bibnamefont {Yamazaki}},\
  }\href@noop {} {\bibfield  {journal} {\bibinfo  {journal} {J. Chem. Phys.}\
  }\textbf {\bibinfo {volume} {100}} (\bibinfo {year} {1994})}\BibitemShut
  {NoStop}%
\bibitem [{\citenamefont {Auhl}\ \emph {et~al.}(2003)\citenamefont {Auhl},
  \citenamefont {Everaers}, \citenamefont {Grest}, \citenamefont {Kremer},\
  and\ \citenamefont {Plimpton}}]{auhl2003equilibration}%
  \BibitemOpen
  \bibfield  {author} {\bibinfo {author} {\bibfnamefont {R.}~\bibnamefont
  {Auhl}}, \bibinfo {author} {\bibfnamefont {R.}~\bibnamefont {Everaers}},
  \bibinfo {author} {\bibfnamefont {G.~S.}\ \bibnamefont {Grest}}, \bibinfo
  {author} {\bibfnamefont {K.}~\bibnamefont {Kremer}}, \ and\ \bibinfo {author}
  {\bibfnamefont {S.~J.}\ \bibnamefont {Plimpton}},\ }\href@noop {} {\bibfield
  {journal} {\bibinfo  {journal} {J. Chem. Phys.}\ }\textbf {\bibinfo {volume}
  {119}},\ \bibinfo {pages} {12718} (\bibinfo {year} {2003})}\BibitemShut
  {NoStop}%
\bibitem [{\citenamefont {Gao}(1995)}]{gao1995efficient}%
  \BibitemOpen
  \bibfield  {author} {\bibinfo {author} {\bibfnamefont {J.}~\bibnamefont
  {Gao}},\ }\href@noop {} {\bibfield  {journal} {\bibinfo  {journal} {J. Chem.
  Phys.}\ }\textbf {\bibinfo {volume} {102}},\ \bibinfo {pages} {1074}
  (\bibinfo {year} {1995})}\BibitemShut {NoStop}%
\bibitem [{\citenamefont {Perez}\ \emph {et~al.}(2008)\citenamefont {Perez},
  \citenamefont {Lame}, \citenamefont {Leonforte},\ and\ \citenamefont
  {Barrat}}]{perez2008polymer}%
  \BibitemOpen
  \bibfield  {author} {\bibinfo {author} {\bibfnamefont {M.}~\bibnamefont
  {Perez}}, \bibinfo {author} {\bibfnamefont {O.}~\bibnamefont {Lame}},
  \bibinfo {author} {\bibfnamefont {F.}~\bibnamefont {Leonforte}}, \ and\
  \bibinfo {author} {\bibfnamefont {J.-L.}\ \bibnamefont {Barrat}},\
  }\href@noop {} {\bibfield  {journal} {\bibinfo  {journal} {J. Chem. Phys.}\
  }\textbf {\bibinfo {volume} {128}},\ \bibinfo {pages} {234904} (\bibinfo
  {year} {2008})}\BibitemShut {NoStop}%
\bibitem [{\citenamefont {Karayiannis}\ \emph
  {et~al.}(2002{\natexlab{a}})\citenamefont {Karayiannis}, \citenamefont
  {Mavrantzas},\ and\ \citenamefont {Theodorou}}]{karayiannis2002novel}%
  \BibitemOpen
  \bibfield  {author} {\bibinfo {author} {\bibfnamefont {N.~C.}\ \bibnamefont
  {Karayiannis}}, \bibinfo {author} {\bibfnamefont {V.~G.}\ \bibnamefont
  {Mavrantzas}}, \ and\ \bibinfo {author} {\bibfnamefont {D.~N.}\ \bibnamefont
  {Theodorou}},\ }\href@noop {} {\bibfield  {journal} {\bibinfo  {journal}
  {Phys. Rev. Lett.}\ }\textbf {\bibinfo {volume} {88}},\ \bibinfo {pages}
  {105503} (\bibinfo {year} {2002}{\natexlab{a}})}\BibitemShut {NoStop}%
\bibitem [{\citenamefont {Mavrantzas}\ \emph {et~al.}(1999)\citenamefont
  {Mavrantzas}, \citenamefont {Boone}, \citenamefont {Zervopoulou},\ and\
  \citenamefont {Theodorou}}]{mavrantzas1999end}%
  \BibitemOpen
  \bibfield  {author} {\bibinfo {author} {\bibfnamefont {V.~G.}\ \bibnamefont
  {Mavrantzas}}, \bibinfo {author} {\bibfnamefont {T.~D.}\ \bibnamefont
  {Boone}}, \bibinfo {author} {\bibfnamefont {E.}~\bibnamefont {Zervopoulou}},
  \ and\ \bibinfo {author} {\bibfnamefont {D.~N.}\ \bibnamefont {Theodorou}},\
  }\href@noop {} {\bibfield  {journal} {\bibinfo  {journal} {Macromolecules}\
  }\textbf {\bibinfo {volume} {32}},\ \bibinfo {pages} {5072} (\bibinfo {year}
  {1999})}\BibitemShut {NoStop}%
\bibitem [{\citenamefont {Pant}\ and\ \citenamefont
  {Theodorou}(1995)}]{pant1995variable}%
  \BibitemOpen
  \bibfield  {author} {\bibinfo {author} {\bibfnamefont {P.~K.}\ \bibnamefont
  {Pant}}\ and\ \bibinfo {author} {\bibfnamefont {D.~N.}\ \bibnamefont
  {Theodorou}},\ }\href@noop {} {\bibfield  {journal} {\bibinfo  {journal}
  {Macromolecules}\ }\textbf {\bibinfo {volume} {28}},\ \bibinfo {pages} {7224}
  (\bibinfo {year} {1995})}\BibitemShut {NoStop}%
\bibitem [{\citenamefont {Karayiannis}\ \emph
  {et~al.}(2002{\natexlab{b}})\citenamefont {Karayiannis}, \citenamefont
  {Giannousaki}, \citenamefont {Mavrantzas},\ and\ \citenamefont
  {Theodorou}}]{karayiannis2002atomistic}%
  \BibitemOpen
  \bibfield  {author} {\bibinfo {author} {\bibfnamefont {N.~C.}\ \bibnamefont
  {Karayiannis}}, \bibinfo {author} {\bibfnamefont {A.~E.}\ \bibnamefont
  {Giannousaki}}, \bibinfo {author} {\bibfnamefont {V.~G.}\ \bibnamefont
  {Mavrantzas}}, \ and\ \bibinfo {author} {\bibfnamefont {D.~N.}\ \bibnamefont
  {Theodorou}},\ }\href@noop {} {\bibfield  {journal} {\bibinfo  {journal} {J.
  Chem. Phys.}\ }\textbf {\bibinfo {volume} {117}},\ \bibinfo {pages} {5465}
  (\bibinfo {year} {2002}{\natexlab{b}})}\BibitemShut {NoStop}%
\bibitem [{\citenamefont {Uhlherr}\ \emph {et~al.}(2002)\citenamefont
  {Uhlherr}, \citenamefont {Doxastakis}, \citenamefont {Mavrantzas},
  \citenamefont {Theodorou}, \citenamefont {Leak}, \citenamefont {Adam},\ and\
  \citenamefont {Nyberg}}]{uhlherr2002atomic}%
  \BibitemOpen
  \bibfield  {author} {\bibinfo {author} {\bibfnamefont {A.}~\bibnamefont
  {Uhlherr}}, \bibinfo {author} {\bibfnamefont {M.}~\bibnamefont {Doxastakis}},
  \bibinfo {author} {\bibfnamefont {V.}~\bibnamefont {Mavrantzas}}, \bibinfo
  {author} {\bibfnamefont {D.}~\bibnamefont {Theodorou}}, \bibinfo {author}
  {\bibfnamefont {S.}~\bibnamefont {Leak}}, \bibinfo {author} {\bibfnamefont
  {N.}~\bibnamefont {Adam}}, \ and\ \bibinfo {author} {\bibfnamefont
  {P.}~\bibnamefont {Nyberg}},\ }\href@noop {} {\bibfield  {journal} {\bibinfo
  {journal} {Europhys. Lett.}\ }\textbf {\bibinfo {volume} {57}},\ \bibinfo
  {pages} {506} (\bibinfo {year} {2002})}\BibitemShut {NoStop}%
\bibitem [{\citenamefont {Karayiannis}\ \emph {et~al.}(2003)\citenamefont
  {Karayiannis}, \citenamefont {Giannousaki},\ and\ \citenamefont
  {Mavrantzas}}]{karayiannis2003advanced}%
  \BibitemOpen
  \bibfield  {author} {\bibinfo {author} {\bibfnamefont {N.~C.}\ \bibnamefont
  {Karayiannis}}, \bibinfo {author} {\bibfnamefont {A.~E.}\ \bibnamefont
  {Giannousaki}}, \ and\ \bibinfo {author} {\bibfnamefont {V.~G.}\ \bibnamefont
  {Mavrantzas}},\ }\href@noop {} {\bibfield  {journal} {\bibinfo  {journal} {J.
  Chem. Phys.}\ }\textbf {\bibinfo {volume} {118}},\ \bibinfo {pages} {2451}
  (\bibinfo {year} {2003})}\BibitemShut {NoStop}%
\bibitem [{\citenamefont {Peristeras}\ \emph {et~al.}(2005)\citenamefont
  {Peristeras}, \citenamefont {Economou},\ and\ \citenamefont
  {Theodorou}}]{peristeras2005structure}%
  \BibitemOpen
  \bibfield  {author} {\bibinfo {author} {\bibfnamefont {L.~D.}\ \bibnamefont
  {Peristeras}}, \bibinfo {author} {\bibfnamefont {I.~G.}\ \bibnamefont
  {Economou}}, \ and\ \bibinfo {author} {\bibfnamefont {D.~N.}\ \bibnamefont
  {Theodorou}},\ }\href@noop {} {\bibfield  {journal} {\bibinfo  {journal}
  {Macromolecules}\ }\textbf {\bibinfo {volume} {38}},\ \bibinfo {pages} {386}
  (\bibinfo {year} {2005})}\BibitemShut {NoStop}%
\bibitem [{\citenamefont {Ramos}\ \emph {et~al.}(2007)\citenamefont {Ramos},
  \citenamefont {Peristeras},\ and\ \citenamefont
  {Theodorou}}]{ramos2007monte}%
  \BibitemOpen
  \bibfield  {author} {\bibinfo {author} {\bibfnamefont {J.}~\bibnamefont
  {Ramos}}, \bibinfo {author} {\bibfnamefont {L.~D.}\ \bibnamefont
  {Peristeras}}, \ and\ \bibinfo {author} {\bibfnamefont {D.~N.}\ \bibnamefont
  {Theodorou}},\ }\href@noop {} {\bibfield  {journal} {\bibinfo  {journal}
  {Macromolecules}\ }\textbf {\bibinfo {volume} {40}},\ \bibinfo {pages} {9640}
  (\bibinfo {year} {2007})}\BibitemShut {NoStop}%
\bibitem [{\citenamefont {Daoulas}\ \emph {et~al.}(2002)\citenamefont
  {Daoulas}, \citenamefont {Terzis},\ and\ \citenamefont
  {Mavrantzas}}]{daoulas2002detailed}%
  \BibitemOpen
  \bibfield  {author} {\bibinfo {author} {\bibfnamefont {K.~C.}\ \bibnamefont
  {Daoulas}}, \bibinfo {author} {\bibfnamefont {A.~F.}\ \bibnamefont {Terzis}},
  \ and\ \bibinfo {author} {\bibfnamefont {V.~G.}\ \bibnamefont {Mavrantzas}},\
  }\href@noop {} {\bibfield  {journal} {\bibinfo  {journal} {J. Chem. Phys.}\
  }\textbf {\bibinfo {volume} {116}},\ \bibinfo {pages} {11028} (\bibinfo
  {year} {2002})}\BibitemShut {NoStop}%
\bibitem [{\citenamefont {Daoulas}\ \emph {et~al.}(2003)\citenamefont
  {Daoulas}, \citenamefont {Terzis},\ and\ \citenamefont
  {Mavrantzas}}]{daoulas2003variable}%
  \BibitemOpen
  \bibfield  {author} {\bibinfo {author} {\bibfnamefont {K.~C.}\ \bibnamefont
  {Daoulas}}, \bibinfo {author} {\bibfnamefont {A.~F.}\ \bibnamefont {Terzis}},
  \ and\ \bibinfo {author} {\bibfnamefont {V.~G.}\ \bibnamefont {Mavrantzas}},\
  }\href@noop {} {\bibfield  {journal} {\bibinfo  {journal} {Macromolecules}\
  }\textbf {\bibinfo {volume} {36}},\ \bibinfo {pages} {6674} (\bibinfo {year}
  {2003})}\BibitemShut {NoStop}%
\bibitem [{\citenamefont {Subramanian}(2010)}]{subramanian2010topology}%
  \BibitemOpen
  \bibfield  {author} {\bibinfo {author} {\bibfnamefont {G.}~\bibnamefont
  {Subramanian}},\ }\href@noop {} {\bibfield  {journal} {\bibinfo  {journal}
  {J. Chem. Phys.}\ }\textbf {\bibinfo {volume} {133}},\ \bibinfo {pages}
  {164902} (\bibinfo {year} {2010})}\BibitemShut {NoStop}%
\bibitem [{\citenamefont {Subramanian}(2011)}]{subramanian2011iterative}%
  \BibitemOpen
  \bibfield  {author} {\bibinfo {author} {\bibfnamefont {G.}~\bibnamefont
  {Subramanian}},\ }\href@noop {} {\bibfield  {journal} {\bibinfo  {journal}
  {Macromol. Theory Simul.}\ }\textbf {\bibinfo {volume} {20}},\ \bibinfo
  {pages} {46} (\bibinfo {year} {2011})}\BibitemShut {NoStop}%
\bibitem [{\citenamefont {Zhang}\ \emph {et~al.}(2014)\citenamefont {Zhang},
  \citenamefont {Moreira}, \citenamefont {Stuehn}, \citenamefont {Daoulas},\
  and\ \citenamefont {Kremer}}]{zhang2014equilibration}%
  \BibitemOpen
  \bibfield  {author} {\bibinfo {author} {\bibfnamefont {G.}~\bibnamefont
  {Zhang}}, \bibinfo {author} {\bibfnamefont {L.~A.}\ \bibnamefont {Moreira}},
  \bibinfo {author} {\bibfnamefont {T.}~\bibnamefont {Stuehn}}, \bibinfo
  {author} {\bibfnamefont {K.~C.}\ \bibnamefont {Daoulas}}, \ and\ \bibinfo
  {author} {\bibfnamefont {K.}~\bibnamefont {Kremer}},\ }\href@noop {}
  {\bibfield  {journal} {\bibinfo  {journal} {ACS Macro. Lett.}\ }\textbf
  {\bibinfo {volume} {3}},\ \bibinfo {pages} {198} (\bibinfo {year}
  {2014})}\BibitemShut {NoStop}%
\bibitem [{\citenamefont {Moreira}\ \emph {et~al.}(2015)\citenamefont
  {Moreira}, \citenamefont {Zhang}, \citenamefont {M{\"u}ller}, \citenamefont
  {Stuehn},\ and\ \citenamefont {Kremer}}]{moreira2015direct}%
  \BibitemOpen
  \bibfield  {author} {\bibinfo {author} {\bibfnamefont {L.~A.}\ \bibnamefont
  {Moreira}}, \bibinfo {author} {\bibfnamefont {G.}~\bibnamefont {Zhang}},
  \bibinfo {author} {\bibfnamefont {F.}~\bibnamefont {M{\"u}ller}}, \bibinfo
  {author} {\bibfnamefont {T.}~\bibnamefont {Stuehn}}, \ and\ \bibinfo {author}
  {\bibfnamefont {K.}~\bibnamefont {Kremer}},\ }\href@noop {} {\bibfield
  {journal} {\bibinfo  {journal} {Macromol. Theory Simul.}\ }\textbf {\bibinfo
  {volume} {24}},\ \bibinfo {pages} {419} (\bibinfo {year} {2015})}\BibitemShut
  {NoStop}%
\bibitem [{\citenamefont {Theodorou}\ and\ \citenamefont
  {Suter}(1985)}]{theodorou1985detailed}%
  \BibitemOpen
  \bibfield  {author} {\bibinfo {author} {\bibfnamefont {D.~N.}\ \bibnamefont
  {Theodorou}}\ and\ \bibinfo {author} {\bibfnamefont {U.~W.}\ \bibnamefont
  {Suter}},\ }\href@noop {} {\bibfield  {journal} {\bibinfo  {journal}
  {Macromolecules}\ }\textbf {\bibinfo {volume} {18}},\ \bibinfo {pages} {1467}
  (\bibinfo {year} {1985})}\BibitemShut {NoStop}%
\bibitem [{\citenamefont {Theodorou}\ and\ \citenamefont
  {Suter}(1986)}]{theodorou1986atomistic}%
  \BibitemOpen
  \bibfield  {author} {\bibinfo {author} {\bibfnamefont {D.~N.}\ \bibnamefont
  {Theodorou}}\ and\ \bibinfo {author} {\bibfnamefont {U.~W.}\ \bibnamefont
  {Suter}},\ }\href@noop {} {\bibfield  {journal} {\bibinfo  {journal}
  {Macromolecules}\ }\textbf {\bibinfo {volume} {19}},\ \bibinfo {pages} {139}
  (\bibinfo {year} {1986})}\BibitemShut {NoStop}%
\bibitem [{\citenamefont {Carbone}\ \emph {et~al.}(2010)\citenamefont
  {Carbone}, \citenamefont {Karimi-Varzaneh},\ and\ \citenamefont
  {M{\"u}ller-Plathe}}]{carbone2010fine}%
  \BibitemOpen
  \bibfield  {author} {\bibinfo {author} {\bibfnamefont {P.}~\bibnamefont
  {Carbone}}, \bibinfo {author} {\bibfnamefont {H.~A.}\ \bibnamefont
  {Karimi-Varzaneh}}, \ and\ \bibinfo {author} {\bibfnamefont {F.}~\bibnamefont
  {M{\"u}ller-Plathe}},\ }\href@noop {} {\bibfield  {journal} {\bibinfo
  {journal} {Faraday Discuss.}\ }\textbf {\bibinfo {volume} {144}},\ \bibinfo
  {pages} {25} (\bibinfo {year} {2010})}\BibitemShut {NoStop}%
\bibitem [{\citenamefont {Kotelyanskii}\ \emph {et~al.}(1996)\citenamefont
  {Kotelyanskii}, \citenamefont {Wagner},\ and\ \citenamefont
  {Paulaitis}}]{kotelyanskii1996building}%
  \BibitemOpen
  \bibfield  {author} {\bibinfo {author} {\bibfnamefont {M.}~\bibnamefont
  {Kotelyanskii}}, \bibinfo {author} {\bibfnamefont {N.}~\bibnamefont
  {Wagner}}, \ and\ \bibinfo {author} {\bibfnamefont {M.~E.}\ \bibnamefont
  {Paulaitis}},\ }\href@noop {} {\bibfield  {journal} {\bibinfo  {journal}
  {Macromolecules}\ }\textbf {\bibinfo {volume} {29}},\ \bibinfo {pages} {8497}
  (\bibinfo {year} {1996})}\BibitemShut {NoStop}%
\bibitem [{\citenamefont {Sliozberg}\ \emph {et~al.}(2016)\citenamefont
  {Sliozberg}, \citenamefont {Kr{\"o}ger},\ and\ \citenamefont
  {Chantawansri}}]{sliozberg2016fast}%
  \BibitemOpen
  \bibfield  {author} {\bibinfo {author} {\bibfnamefont {Y.~R.}\ \bibnamefont
  {Sliozberg}}, \bibinfo {author} {\bibfnamefont {M.}~\bibnamefont
  {Kr{\"o}ger}}, \ and\ \bibinfo {author} {\bibfnamefont {T.~L.}\ \bibnamefont
  {Chantawansri}},\ }\href@noop {} {\bibfield  {journal} {\bibinfo  {journal}
  {The Journal of Chemical Physics}\ }\textbf {\bibinfo {volume} {144}},\
  \bibinfo {pages} {154901} (\bibinfo {year} {2016})}\BibitemShut {NoStop}%
\bibitem [{\citenamefont {Grest}\ and\ \citenamefont
  {Kremer}(1986)}]{grest1986molecular}%
  \BibitemOpen
  \bibfield  {author} {\bibinfo {author} {\bibfnamefont {G.~S.}\ \bibnamefont
  {Grest}}\ and\ \bibinfo {author} {\bibfnamefont {K.}~\bibnamefont {Kremer}},\
  }\href@noop {} {\bibfield  {journal} {\bibinfo  {journal} {Phys. Rev. A}\
  }\textbf {\bibinfo {volume} {33}},\ \bibinfo {pages} {3628} (\bibinfo {year}
  {1986})}\BibitemShut {NoStop}%
\bibitem [{\citenamefont {Svaneborg}\ \emph {et~al.}(2016)\citenamefont
  {Svaneborg}, \citenamefont {Karimi-Varzaneh}, \citenamefont {Hojdis},\ and\
  \citenamefont {Everaers}}]{svaneborgContiII}%
  \BibitemOpen
  \bibfield  {author} {\bibinfo {author} {\bibfnamefont {C.}~\bibnamefont
  {Svaneborg}}, \bibinfo {author} {\bibfnamefont {H.}~\bibnamefont
  {Karimi-Varzaneh}}, \bibinfo {author} {\bibfnamefont {N.}~\bibnamefont
  {Hojdis}}, \ and\ \bibinfo {author} {\bibfnamefont {R.}~\bibnamefont
  {Everaers}},\ }\href@noop {} {\enquote {\bibinfo {title} {Kremer-{G}rest
  models of common polymer melts. {M}anuscript submitted.}}\ } (\bibinfo {year}
  {2016})\BibitemShut {NoStop}%
\bibitem [{\citenamefont {Hoogerbrugge}\ and\ \citenamefont
  {Koelman}(1992)}]{hoogerbrugge1992simulating}%
  \BibitemOpen
  \bibfield  {author} {\bibinfo {author} {\bibfnamefont {P.}~\bibnamefont
  {Hoogerbrugge}}\ and\ \bibinfo {author} {\bibfnamefont {J.}~\bibnamefont
  {Koelman}},\ }\href@noop {} {\bibfield  {journal} {\bibinfo  {journal}
  {Europhys. Lett.}\ }\textbf {\bibinfo {volume} {19}},\ \bibinfo {pages} {155}
  (\bibinfo {year} {1992})}\BibitemShut {NoStop}%
\bibitem [{\citenamefont {Espanol}\ and\ \citenamefont
  {Warren}(1995)}]{espanol1995statistical}%
  \BibitemOpen
  \bibfield  {author} {\bibinfo {author} {\bibfnamefont {P.}~\bibnamefont
  {Espanol}}\ and\ \bibinfo {author} {\bibfnamefont {P.}~\bibnamefont
  {Warren}},\ }\href@noop {} {\bibfield  {journal} {\bibinfo  {journal}
  {Europhys. Lett.}\ }\textbf {\bibinfo {volume} {30}},\ \bibinfo {pages} {191}
  (\bibinfo {year} {1995})}\BibitemShut {NoStop}%
\bibitem [{\citenamefont {Flory}(1949)}]{flory49}%
  \BibitemOpen
  \bibfield  {author} {\bibinfo {author} {\bibfnamefont {P.~J.}\ \bibnamefont
  {Flory}},\ }\href@noop {} {\bibfield  {journal} {\bibinfo  {journal} {J.
  Chem. Phys.}\ }\textbf {\bibinfo {volume} {17}},\ \bibinfo {pages} {303}
  (\bibinfo {year} {1949})}\BibitemShut {NoStop}%
\bibitem [{\citenamefont {Flory}\ and\ \citenamefont
  {Jackson}(1989)}]{flory1989statistical}%
  \BibitemOpen
  \bibfield  {author} {\bibinfo {author} {\bibfnamefont {P.}~\bibnamefont
  {Flory}}\ and\ \bibinfo {author} {\bibfnamefont {J.}~\bibnamefont
  {Jackson}},\ }\href@noop {} {\emph {\bibinfo {title} {Statistical Mechanics
  of Chain Molecules}}}\ (\bibinfo  {publisher} {Hanser},\ \bibinfo {year}
  {1989})\BibitemShut {NoStop}%
\bibitem [{\citenamefont {Edwards}(1967)}]{edwards67}%
  \BibitemOpen
  \bibfield  {author} {\bibinfo {author} {\bibfnamefont {S.}~\bibnamefont
  {Edwards}},\ }\href@noop {} {\ \textbf {\bibinfo {volume} {91}},\ \bibinfo
  {pages} {513} (\bibinfo {year} {1967})}\BibitemShut {NoStop}%
\bibitem [{\citenamefont {Rouse~Jr}(1953)}]{rouse1953theory}%
  \BibitemOpen
  \bibfield  {author} {\bibinfo {author} {\bibfnamefont {P.~E.}\ \bibnamefont
  {Rouse~Jr}},\ }\href@noop {} {\bibfield  {journal} {\bibinfo  {journal} {J.
  Chem. Phys.}\ }\textbf {\bibinfo {volume} {21}},\ \bibinfo {pages} {1272}
  (\bibinfo {year} {1953})}\BibitemShut {NoStop}%
\bibitem [{\citenamefont {Everaers}(2012)}]{everaers2012topological}%
  \BibitemOpen
  \bibfield  {author} {\bibinfo {author} {\bibfnamefont {R.}~\bibnamefont
  {Everaers}},\ }\href@noop {} {\bibfield  {journal} {\bibinfo  {journal}
  {Phys. Rev. E.}\ }\textbf {\bibinfo {volume} {86}},\ \bibinfo {pages}
  {022801} (\bibinfo {year} {2012})}\BibitemShut {NoStop}%
\bibitem [{\citenamefont {Kremer}\ and\ \citenamefont
  {Grest}(1990)}]{kremer1990dynamics}%
  \BibitemOpen
  \bibfield  {author} {\bibinfo {author} {\bibfnamefont {K.}~\bibnamefont
  {Kremer}}\ and\ \bibinfo {author} {\bibfnamefont {G.~S.}\ \bibnamefont
  {Grest}},\ }\href@noop {} {\bibfield  {journal} {\bibinfo  {journal} {J.
  Chem. Phys.}\ }\textbf {\bibinfo {volume} {92}},\ \bibinfo {pages} {5057}
  (\bibinfo {year} {1990})}\BibitemShut {NoStop}%
\bibitem [{\citenamefont {Faller}\ \emph {et~al.}(1999)\citenamefont {Faller},
  \citenamefont {Kolb},\ and\ \citenamefont
  {M{\"u}ller-Plathe}}]{faller1999local}%
  \BibitemOpen
  \bibfield  {author} {\bibinfo {author} {\bibfnamefont {R.}~\bibnamefont
  {Faller}}, \bibinfo {author} {\bibfnamefont {A.}~\bibnamefont {Kolb}}, \ and\
  \bibinfo {author} {\bibfnamefont {F.}~\bibnamefont {M{\"u}ller-Plathe}},\
  }\href@noop {} {\bibfield  {journal} {\bibinfo  {journal} {Phys. Chem. Chem.
  Phys.}\ }\textbf {\bibinfo {volume} {1}},\ \bibinfo {pages} {2071} (\bibinfo
  {year} {1999})}\BibitemShut {NoStop}%
\bibitem [{\citenamefont {Faller}\ \emph {et~al.}(2000)\citenamefont {Faller},
  \citenamefont {M{\"u}ller-Plathe},\ and\ \citenamefont
  {Heuer}}]{faller2000local}%
  \BibitemOpen
  \bibfield  {author} {\bibinfo {author} {\bibfnamefont {R.}~\bibnamefont
  {Faller}}, \bibinfo {author} {\bibfnamefont {F.}~\bibnamefont
  {M{\"u}ller-Plathe}}, \ and\ \bibinfo {author} {\bibfnamefont
  {A.}~\bibnamefont {Heuer}},\ }\href@noop {} {\bibfield  {journal} {\bibinfo
  {journal} {Macromolecules}\ }\textbf {\bibinfo {volume} {33}},\ \bibinfo
  {pages} {6602} (\bibinfo {year} {2000})}\BibitemShut {NoStop}%
\bibitem [{\citenamefont {Faller}\ and\ \citenamefont
  {M{\"u}ller-Plathe}(2001)}]{faller2001chain}%
  \BibitemOpen
  \bibfield  {author} {\bibinfo {author} {\bibfnamefont {R.}~\bibnamefont
  {Faller}}\ and\ \bibinfo {author} {\bibfnamefont {F.}~\bibnamefont
  {M{\"u}ller-Plathe}},\ }\href@noop {} {\bibfield  {journal} {\bibinfo
  {journal} {Chem. Phys. Chem.}\ }\textbf {\bibinfo {volume} {2}},\ \bibinfo
  {pages} {180} (\bibinfo {year} {2001})}\BibitemShut {NoStop}%
\bibitem [{\citenamefont {Gr{\o}nbech-Jensen}\ and\ \citenamefont
  {Farago}(2013)}]{gronbech2013simple}%
  \BibitemOpen
  \bibfield  {author} {\bibinfo {author} {\bibfnamefont {N.}~\bibnamefont
  {Gr{\o}nbech-Jensen}}\ and\ \bibinfo {author} {\bibfnamefont
  {O.}~\bibnamefont {Farago}},\ }\href@noop {} {\bibfield  {journal} {\bibinfo
  {journal} {Mol. Phys.}\ }\textbf {\bibinfo {volume} {111}},\ \bibinfo {pages}
  {983} (\bibinfo {year} {2013})}\BibitemShut {NoStop}%
\bibitem [{\citenamefont {Gr{\o}nbech-Jensen}\ \emph
  {et~al.}(2014)\citenamefont {Gr{\o}nbech-Jensen}, \citenamefont {Hayre},\
  and\ \citenamefont {Farago}}]{gronbech2014application}%
  \BibitemOpen
  \bibfield  {author} {\bibinfo {author} {\bibfnamefont {N.}~\bibnamefont
  {Gr{\o}nbech-Jensen}}, \bibinfo {author} {\bibfnamefont {N.~R.}\ \bibnamefont
  {Hayre}}, \ and\ \bibinfo {author} {\bibfnamefont {O.}~\bibnamefont
  {Farago}},\ }\href@noop {} {\bibfield  {journal} {\bibinfo  {journal} {Comp.
  Phys. Comm.}\ }\textbf {\bibinfo {volume} {185}},\ \bibinfo {pages} {524}
  (\bibinfo {year} {2014})}\BibitemShut {NoStop}%
\bibitem [{\citenamefont {Plimpton}(1995)}]{PlimptonLAMMPS}%
  \BibitemOpen
  \bibfield  {author} {\bibinfo {author} {\bibfnamefont {S.}~\bibnamefont
  {Plimpton}},\ }\href@noop {} {\bibfield  {journal} {\bibinfo  {journal} {J.
  Comp. Phys.}\ }\textbf {\bibinfo {volume} {117}},\ \bibinfo {pages} {1}
  (\bibinfo {year} {1995})}\BibitemShut {NoStop}%
\bibitem [{\citenamefont {Sukumaran}\ \emph {et~al.}(2005)\citenamefont
  {Sukumaran}, \citenamefont {Grest}, \citenamefont {Kremer},\ and\
  \citenamefont {Everaers}}]{sukumaran2005identifying}%
  \BibitemOpen
  \bibfield  {author} {\bibinfo {author} {\bibfnamefont {S.~K.}\ \bibnamefont
  {Sukumaran}}, \bibinfo {author} {\bibfnamefont {G.~S.}\ \bibnamefont
  {Grest}}, \bibinfo {author} {\bibfnamefont {K.}~\bibnamefont {Kremer}}, \
  and\ \bibinfo {author} {\bibfnamefont {R.}~\bibnamefont {Everaers}},\
  }\href@noop {} {\bibfield  {journal} {\bibinfo  {journal} {J. Polym. Sci.,
  Part B: Polym. Phys.}\ }\textbf {\bibinfo {volume} {43}},\ \bibinfo {pages}
  {917} (\bibinfo {year} {2005})}\BibitemShut {NoStop}%
\bibitem [{\citenamefont {Fetters}\ \emph {et~al.}(2007)\citenamefont
  {Fetters}, \citenamefont {Lohse},\ and\ \citenamefont
  {Colby}}]{fetters2007chain}%
  \BibitemOpen
  \bibfield  {author} {\bibinfo {author} {\bibfnamefont {L.}~\bibnamefont
  {Fetters}}, \bibinfo {author} {\bibfnamefont {D.}~\bibnamefont {Lohse}}, \
  and\ \bibinfo {author} {\bibfnamefont {R.}~\bibnamefont {Colby}},\ }in\
  \href@noop {} {\emph {\bibinfo {booktitle} {Physical properties of polymers
  handbook}}},\ \bibinfo {editor} {edited by\ \bibinfo {editor} {\bibfnamefont
  {J.}~\bibnamefont {Mark}}}\ (\bibinfo  {publisher} {Springer},\ \bibinfo
  {year} {2007})\ pp.\ \bibinfo {pages} {447--454}\BibitemShut {NoStop}%
\bibitem [{\citenamefont {Wang}(2009)}]{wang2009studying}%
  \BibitemOpen
  \bibfield  {author} {\bibinfo {author} {\bibfnamefont {Q.}~\bibnamefont
  {Wang}},\ }\href@noop {} {\bibfield  {journal} {\bibinfo  {journal} {Soft
  Matter}\ }\textbf {\bibinfo {volume} {5}},\ \bibinfo {pages} {4564} (\bibinfo
  {year} {2009})}\BibitemShut {NoStop}%
\bibitem [{\citenamefont {Helfand}\ and\ \citenamefont
  {Tagami}(1971)}]{helfand1971theory}%
  \BibitemOpen
  \bibfield  {author} {\bibinfo {author} {\bibfnamefont {E.}~\bibnamefont
  {Helfand}}\ and\ \bibinfo {author} {\bibfnamefont {Y.}~\bibnamefont
  {Tagami}},\ }\href@noop {} {\bibfield  {journal} {\bibinfo  {journal} {J.
  Polym. Sci. B.}\ }\textbf {\bibinfo {volume} {9}},\ \bibinfo {pages} {741}
  (\bibinfo {year} {1971})}\BibitemShut {NoStop}%
\bibitem [{\citenamefont {Helfand}\ and\ \citenamefont
  {Tagami}(1972)}]{helfand1972theory}%
  \BibitemOpen
  \bibfield  {author} {\bibinfo {author} {\bibfnamefont {E.}~\bibnamefont
  {Helfand}}\ and\ \bibinfo {author} {\bibfnamefont {Y.}~\bibnamefont
  {Tagami}},\ }\href@noop {} {\bibfield  {journal} {\bibinfo  {journal} {J.
  Chem. Phys.}\ }\textbf {\bibinfo {volume} {56}},\ \bibinfo {pages} {3592}
  (\bibinfo {year} {1972})}\BibitemShut {NoStop}%
\bibitem [{\citenamefont {Madras}\ and\ \citenamefont
  {Sokal}(1988)}]{madras1988pivot}%
  \BibitemOpen
  \bibfield  {author} {\bibinfo {author} {\bibfnamefont {N.}~\bibnamefont
  {Madras}}\ and\ \bibinfo {author} {\bibfnamefont {A.~D.}\ \bibnamefont
  {Sokal}},\ }\href@noop {} {\bibfield  {journal} {\bibinfo  {journal} {J.
  Stat. Phys.}\ }\textbf {\bibinfo {volume} {50}},\ \bibinfo {pages} {109}
  (\bibinfo {year} {1988})}\BibitemShut {NoStop}%
\bibitem [{\citenamefont {Wittmer}\ \emph
  {et~al.}(2007{\natexlab{a}})\citenamefont {Wittmer}, \citenamefont
  {Beckrich}, \citenamefont {Johner}, \citenamefont {Semenov}, \citenamefont
  {Obukhov}, \citenamefont {Meyer},\ and\ \citenamefont
  {Baschnagel}}]{wittmer2007polymer}%
  \BibitemOpen
  \bibfield  {author} {\bibinfo {author} {\bibfnamefont {J.}~\bibnamefont
  {Wittmer}}, \bibinfo {author} {\bibfnamefont {P.}~\bibnamefont {Beckrich}},
  \bibinfo {author} {\bibfnamefont {A.}~\bibnamefont {Johner}}, \bibinfo
  {author} {\bibfnamefont {A.}~\bibnamefont {Semenov}}, \bibinfo {author}
  {\bibfnamefont {S.}~\bibnamefont {Obukhov}}, \bibinfo {author} {\bibfnamefont
  {H.}~\bibnamefont {Meyer}}, \ and\ \bibinfo {author} {\bibfnamefont
  {J.}~\bibnamefont {Baschnagel}},\ }\href@noop {} {\bibfield  {journal}
  {\bibinfo  {journal} {Europhys. Lett.}\ }\textbf {\bibinfo {volume} {77}},\
  \bibinfo {pages} {56003} (\bibinfo {year} {2007}{\natexlab{a}})}\BibitemShut
  {NoStop}%
\bibitem [{\citenamefont {Wittmer}\ \emph
  {et~al.}(2007{\natexlab{b}})\citenamefont {Wittmer}, \citenamefont
  {Beckrich}, \citenamefont {Meyer}, \citenamefont {Cavallo}, \citenamefont
  {Johner},\ and\ \citenamefont {Baschnagel}}]{wittmer2007intramolecular}%
  \BibitemOpen
  \bibfield  {author} {\bibinfo {author} {\bibfnamefont {J.}~\bibnamefont
  {Wittmer}}, \bibinfo {author} {\bibfnamefont {P.}~\bibnamefont {Beckrich}},
  \bibinfo {author} {\bibfnamefont {H.}~\bibnamefont {Meyer}}, \bibinfo
  {author} {\bibfnamefont {A.}~\bibnamefont {Cavallo}}, \bibinfo {author}
  {\bibfnamefont {A.}~\bibnamefont {Johner}}, \ and\ \bibinfo {author}
  {\bibfnamefont {J.}~\bibnamefont {Baschnagel}},\ }\href@noop {} {\bibfield
  {journal} {\bibinfo  {journal} {Phys. Rev. E.}\ }\textbf {\bibinfo {volume}
  {76}},\ \bibinfo {pages} {011803} (\bibinfo {year}
  {2007}{\natexlab{b}})}\BibitemShut {NoStop}%
\bibitem [{\citenamefont {Beckrich}\ \emph {et~al.}(2007)\citenamefont
  {Beckrich}, \citenamefont {Johner}, \citenamefont {Semenov}, \citenamefont
  {Obukhov}, \citenamefont {Benoit},\ and\ \citenamefont
  {Wittmer}}]{beckrich2007intramolecular}%
  \BibitemOpen
  \bibfield  {author} {\bibinfo {author} {\bibfnamefont {P.}~\bibnamefont
  {Beckrich}}, \bibinfo {author} {\bibfnamefont {A.}~\bibnamefont {Johner}},
  \bibinfo {author} {\bibfnamefont {A.~N.}\ \bibnamefont {Semenov}}, \bibinfo
  {author} {\bibfnamefont {S.~P.}\ \bibnamefont {Obukhov}}, \bibinfo {author}
  {\bibfnamefont {H.}~\bibnamefont {Benoit}}, \ and\ \bibinfo {author}
  {\bibfnamefont {J.}~\bibnamefont {Wittmer}},\ }\href@noop {} {\bibfield
  {journal} {\bibinfo  {journal} {Macromolecules}\ }\textbf {\bibinfo {volume}
  {40}},\ \bibinfo {pages} {3805} (\bibinfo {year} {2007})}\BibitemShut
  {NoStop}%
\bibitem [{\citenamefont {Meyer}\ \emph {et~al.}(2008)\citenamefont {Meyer},
  \citenamefont {Wittmer}, \citenamefont {Kreer}, \citenamefont {Beckrich},
  \citenamefont {Johner}, \citenamefont {Farago},\ and\ \citenamefont
  {Baschnagel}}]{meyer2008static}%
  \BibitemOpen
  \bibfield  {author} {\bibinfo {author} {\bibfnamefont {H.}~\bibnamefont
  {Meyer}}, \bibinfo {author} {\bibfnamefont {J.}~\bibnamefont {Wittmer}},
  \bibinfo {author} {\bibfnamefont {T.}~\bibnamefont {Kreer}}, \bibinfo
  {author} {\bibfnamefont {P.}~\bibnamefont {Beckrich}}, \bibinfo {author}
  {\bibfnamefont {A.}~\bibnamefont {Johner}}, \bibinfo {author} {\bibfnamefont
  {J.}~\bibnamefont {Farago}}, \ and\ \bibinfo {author} {\bibfnamefont
  {J.}~\bibnamefont {Baschnagel}},\ }\href@noop {} {\bibfield  {journal}
  {\bibinfo  {journal} {Eur. Phys. J. E.}\ }\textbf {\bibinfo {volume} {26}},\
  \bibinfo {pages} {25} (\bibinfo {year} {2008})}\BibitemShut {NoStop}%
\bibitem [{\citenamefont {Semenov}(2010)}]{semenov2010bond}%
  \BibitemOpen
  \bibfield  {author} {\bibinfo {author} {\bibfnamefont {A.}~\bibnamefont
  {Semenov}},\ }\href@noop {} {\bibfield  {journal} {\bibinfo  {journal}
  {Macromolecules}\ }\textbf {\bibinfo {volume} {43}},\ \bibinfo {pages} {9139}
  (\bibinfo {year} {2010})}\BibitemShut {NoStop}%
\bibitem [{\citenamefont {Sides}\ \emph {et~al.}(2004)\citenamefont {Sides},
  \citenamefont {Grest}, \citenamefont {Stevens},\ and\ \citenamefont
  {Plimpton}}]{sides2004effect}%
  \BibitemOpen
  \bibfield  {author} {\bibinfo {author} {\bibfnamefont {S.~W.}\ \bibnamefont
  {Sides}}, \bibinfo {author} {\bibfnamefont {G.~S.}\ \bibnamefont {Grest}},
  \bibinfo {author} {\bibfnamefont {M.~J.}\ \bibnamefont {Stevens}}, \ and\
  \bibinfo {author} {\bibfnamefont {S.~J.}\ \bibnamefont {Plimpton}},\
  }\href@noop {} {\bibfield  {journal} {\bibinfo  {journal} {J. Polym. Sci.,
  Part B: Polym. Phys.}\ }\textbf {\bibinfo {volume} {42}},\ \bibinfo {pages}
  {199} (\bibinfo {year} {2004})}\BibitemShut {NoStop}%
\bibitem [{\citenamefont {Hoy}\ \emph {et~al.}(2009)\citenamefont {Hoy},
  \citenamefont {Foteinopoulou},\ and\ \citenamefont {Kr\"{o}ger}}]{HoyPRE09}%
  \BibitemOpen
  \bibfield  {author} {\bibinfo {author} {\bibfnamefont {R.~S.}\ \bibnamefont
  {Hoy}}, \bibinfo {author} {\bibfnamefont {K.}~\bibnamefont {Foteinopoulou}},
  \ and\ \bibinfo {author} {\bibfnamefont {M.}~\bibnamefont {Kr\"{o}ger}},\
  }\href@noop {} {\bibfield  {journal} {\bibinfo  {journal} {Phys. Rev. E}\
  }\textbf {\bibinfo {volume} {80}},\ \bibinfo {pages} {031803} (\bibinfo
  {year} {2009})}\BibitemShut {NoStop}%
\bibitem [{\citenamefont {Zhang}\ \emph {et~al.}(2015)\citenamefont {Zhang},
  \citenamefont {Stuehn}, \citenamefont {Daoulas},\ and\ \citenamefont
  {Kremer}}]{zhang2015communication}%
  \BibitemOpen
  \bibfield  {author} {\bibinfo {author} {\bibfnamefont {G.}~\bibnamefont
  {Zhang}}, \bibinfo {author} {\bibfnamefont {T.}~\bibnamefont {Stuehn}},
  \bibinfo {author} {\bibfnamefont {K.~C.}\ \bibnamefont {Daoulas}}, \ and\
  \bibinfo {author} {\bibfnamefont {K.}~\bibnamefont {Kremer}},\ }\href@noop {}
  {\bibfield  {journal} {\bibinfo  {journal} {J. Chem. Phys.}\ }\textbf
  {\bibinfo {volume} {142}},\ \bibinfo {pages} {221102} (\bibinfo {year}
  {2015})}\BibitemShut {NoStop}%
\bibitem [{\citenamefont {Hoy}\ and\ \citenamefont
  {Robbins}(2005)}]{hoy2005eep}%
  \BibitemOpen
  \bibfield  {author} {\bibinfo {author} {\bibfnamefont {R.~S.}\ \bibnamefont
  {Hoy}}\ and\ \bibinfo {author} {\bibfnamefont {M.~O.}\ \bibnamefont
  {Robbins}},\ }\href@noop {} {\bibfield  {journal} {\bibinfo  {journal}
  {Physical Review E}\ }\textbf {\bibinfo {volume} {72}},\ \bibinfo {pages}
  {61802} (\bibinfo {year} {2005})}\BibitemShut {NoStop}%
\bibitem [{Aba()}]{AbacusHardware}%
  \BibitemOpen
  \href@noop {} {\enquote {\bibinfo {title} {Hardware setup of the abacus 2.0
  cluster at the university of southern denmark.
  https://deic.sdu.dk/setup/hardware},}\ }\BibitemShut {NoStop}%
\bibitem [{\citenamefont {Svaneborg}\ and\ \citenamefont
  {Pedersen}(2012{\natexlab{a}})}]{svaneborg2012formalism}%
  \BibitemOpen
  \bibfield  {author} {\bibinfo {author} {\bibfnamefont {C.}~\bibnamefont
  {Svaneborg}}\ and\ \bibinfo {author} {\bibfnamefont {J.~S.}\ \bibnamefont
  {Pedersen}},\ }\href@noop {} {\bibfield  {journal} {\bibinfo  {journal} {J.
  Chem. Phys.}\ }\textbf {\bibinfo {volume} {136}},\ \bibinfo {pages} {104105}
  (\bibinfo {year} {2012}{\natexlab{a}})}\BibitemShut {NoStop}%
\bibitem [{\citenamefont {Svaneborg}\ and\ \citenamefont
  {Pedersen}(2012{\natexlab{b}})}]{svaneborg2012formalism2}%
  \BibitemOpen
  \bibfield  {author} {\bibinfo {author} {\bibfnamefont {C.}~\bibnamefont
  {Svaneborg}}\ and\ \bibinfo {author} {\bibfnamefont {J.~S.}\ \bibnamefont
  {Pedersen}},\ }\href@noop {} {\bibfield  {journal} {\bibinfo  {journal} {J.
  Chem. Phys.}\ }\textbf {\bibinfo {volume} {136}},\ \bibinfo {pages} {154907}
  (\bibinfo {year} {2012}{\natexlab{b}})}\BibitemShut {NoStop}%
\bibitem [{\citenamefont {Debye}(1947)}]{debye1947formfactor}%
  \BibitemOpen
  \bibfield  {author} {\bibinfo {author} {\bibfnamefont {P.}~\bibnamefont
  {Debye}},\ }\href@noop {} {\bibfield  {journal} {\bibinfo  {journal} {J.
  Phys. Colloid Chem.}\ }\textbf {\bibinfo {volume} {51}},\ \bibinfo {pages}
  {18} (\bibinfo {year} {1947})}\BibitemShut {NoStop}%
\bibitem [{\citenamefont {Hammouda}(1992)}]{hammouda1992formfactoramplitude}%
  \BibitemOpen
  \bibfield  {author} {\bibinfo {author} {\bibfnamefont {B.}~\bibnamefont
  {Hammouda}},\ }\href@noop {} {\bibfield  {journal} {\bibinfo  {journal} {J.
  Polym. Sci., Part B: Polym. Phys.}\ }\textbf {\bibinfo {volume} {30}},\
  \bibinfo {pages} {1387} (\bibinfo {year} {1992})}\BibitemShut {NoStop}%
\end{thebibliography}

%

\end{document}